\documentclass[aps,pre,reprint,amsmath,superscriptaddress,showpacs,floatfix]{revtex4-1}
\usepackage{graphicx,color,hyperref,siunitx}

\begin{document}

\title{Disentangling synchrony from serial dependency in paired event time series}

\author{Adrian Odenweller}
\email{adrian.odenweller@pik-potsdam.de}
\affiliation{Potsdam Institute for Climate Impact Research (PIK), Germany}
\affiliation{Center for Earth System Research and Sustainability (CEN), University of Hamburg, Germany}
\affiliation{The Land in the Earth System, Max Planck Institute for Meteorology, Hamburg, Germany}

\author{Reik V. Donner}
\email{reik.donner@pik-potsdam.de, reik.donner@h2.de}
\affiliation{Potsdam Institute for Climate Impact Research (PIK), Germany}
\affiliation{Department of Water, Environment, Construction and Safety, Magdeburg--Stendal University of Applied Sciences, Magdeburg, Germany}

\date{\today}

\begin{abstract}
Quantifying synchronization phenomena based on the timing of events has recently attracted a great deal of interest in various disciplines such as neuroscience or climatology. A multitude of similarity measures has been proposed for this purpose, including Event Synchronization (ES) and Event Coincidence Analysis (ECA) as two widely applicable examples. While ES defines synchrony in a data adaptive local way that does not distinguish between different time scales, ECA requires selecting a specific scale for analysis. In this paper, we use slightly modified versions of both ES and ECA that address previous issues with respect to proper normalization and boundary treatment, which are particularly relevant for short time series with low temporal resolution. By numerically studying threshold crossing events in coupled autoregressive processes, we identify a practical limitation of ES when attempting to study synchrony between serially dependent event sequences exhibiting event clustering in time. Practical implications of this observation are demonstrated for the case of functional network representations of climate extremes based on both ES and ECA, while no marked differences between both measures are observed for the case of epileptic electroencephalogram (EEG) data. Our findings suggest that careful event detection along with diligent preprocessing is recommended when applying ES while less crucial for ECA. Despite the lack of a general \emph{modus operandi} for both event definition and detection of synchronization, we suggest ECA as a widely robust method, especially for time resolved synchronization analyses of event time series from various disciplines.
\end{abstract}

\pacs{05.45.Tp, 05.45.Xt, 92.60.Ry, 87.19.La}

\maketitle

\section{Introduction \label{sec:intro}}

Along with the rising availability of empirical data across many scientific fields, in the past decades a variety of statistical methods have been newly developed to deal with ever larger datasets. As specific events, particularly extremes, in both nature and society attract a great deal of attention from the academic world as well as the general public \cite{Albeverio2006}, a methodologically sound analysis of event time series in general, and synchrony between event in particular, is crucial not only for research progress, but also for informed decision making relying on confident results. Accordingly, in different fields of science, the quantification of event synchronization has recently become a focal point of a plethora of different studies and methods. Among the methodological developments aimed at serving this purpose, event synchronization (ES) \cite{QuianQuiroga2002} and event coincidence analysis (ECA) \cite{Donges2011,Donges2016} stand out as two conceptually simple nonlinear measures that are potentially applicable to a broad variety of problems of such diverse fields like neuroscience and climatology.

Being originally motivated by the emerging nonlinear dynamical analysis of electroencephalogram (EEG) recordings in terms of spike train synchrony between different brain areas \cite{QuianQuiroga2002, Kreuz2004, Pereda2005, Kreuz2007, Kreuz2007a, Stam2005, Dauwels2008, Angotzi2014, Singh2014}, ES has recently been applied to problems outside the neurosciences as well, including group dynamics in both humans and animals \cite{Varni2010, Butail2016}, econophysics \cite{Zhou2003}, and climate extremes \cite{Malik2010, Malik2012, Boers2013, Boers2014, Boers2014a, Boers2015, Boers2015a, Boers2016, Stolbova2014, He2014, Rehfeld2014, Rheinwalt2012, Marwan2015}. Notably, the thorough application to climate problems has been mainly governed by a methodological combination with the paradigm of functional network analysis \cite{Donner2017,Dijkstra2019}, as will be further detailed in the course of this work. ES has the important advantage of automatically classifying pairs of events at two distinct spatial locations as (not) synchronized without the need to manually select any algorithmic parameters, particularly a maximum tolerable mutual delay to consider two events synchronized.

On the other hand, ECA has been recently introduced based on the general idea of capturing event synchrony between point processes, which do not necessarily share the common properties of neuronal spike trains like a relatively well expressed pacemaker. Successful applications of the method can be found across various disciplines, including paleoclimatology \cite{Donges2011}, the climate-security nexus \cite{Donges2016, Schleussner2016}, plant sciences \cite{Siegmund2016, Siegmund2016a, Siegmund2017, Rammig2015, Baumbach2017}, modern day climatology \cite{Wiedermann2017, Sippel2017} and even seismology \cite{Sarlis2018}. As opposed to ES, ECA commonly requires at least one input parameter (the maximum possible delay) to be selected, which however also entails the potential advantage of a more refined analysis by isolating certain time scales based on a priori knowledge or specific research questions. Unlike ES, ECA has so far not been used to analyze EEG data or generate functional network representations of large scale spatiotemporal climate data, making this a novelty worth exploring in the forthcoming sections.

Both ES and ECA share the fundamental property of basically counting synchronous events based on pairwise comparison and subsequent aggregation. Yet, they differ in the specification of the tolerance window for identifying synchronous events, with ES relying on a dynamic (data adaptive) and hence local approach, while ECA requests a static (global) parameter to be selected. Even though both methods have been demonstrated to be applicable to a wide range of research problems, they have exhibited a tendency to remain used by rather disjoint scientific communities, as no in-depth comparison, which elaborates on the (dis)advantages of each method, has been conducted so far.

Accordingly, in the present paper, we attempt to provide a thorough comparison between ES and ECA by means of numerical simulation of simple coupled stochastic processes along with real world applications to two exemplary climate and EEG datasets. Firstly, in Section \ref{sec:method}, we introduce formally correct variants of both association measures that address previous ambiguities in the counting procedures, thereby re-establishing proper normalization, which is especially relevant for short time series with low temporal resolution. In Section \ref{sec:artificial}, we consider coupled autoregressive AR(1) processes to demonstrate that ES has structural difficulties to capture synchrony in the case of events that are temporally clustered, i.e., serially dependent.

This clustering is no typical feature of EEG recordings but emerges rather commonly in climate datasets \cite{Wolf2020}, as we illustrate with a functional network analysis of the Indian monsoon system that replicates previous studies and highlights their methodological deficiencies (Section \ref{subsec:climdata}). Therefore, we argue that previous research results need to be interpreted with caution. On the contrary, we provide evidence that ECA is not markedly affected by event clustering, but provides the additional benefit of allowing for testing physical hypotheses via systematically varying the associated parameter settings (which may be guided by a priori knowledge of the system). Building on these conceptual concerns, we propose ECA as a promising robust alternative to ES if temporal event clustering cannot be ruled out. By analyzing epileptic rat EEG signals (Section \ref{subsec:eeg}), we find that for time series with relatively regularly spaced peaks, ES and ECA are practically interchangeable. However, if the integration of different time scales into one measure is contextually justified and events can be clearly marked-off, ES may still be the favorable method, since it does not require any parameter selection by the user. In Section \ref{sec:discussion}, we discuss the repercussions of event definitions on the choice of association measures that can be derived from our numerical results. Finally, some general conclusions are drawn in Section \ref{sec:conclusion}.

\section{Methods \label{sec:method}}

Both ES and ECA provide measures that go beyond second-order moments captured by classical correlations. They are based on pairs of event sequences or binary event time series as inputs. When being initially given two ``normal'' time series of nonbinary (continuous or discrete valued) variables, such sequences are often obtained by applying a threshold to the underlying time series at a given percentile (other options will be discussed later). We extract the time indices of these threshold exceedance events and let $t_l^i$ denote the time of event $l$ in time series $i$ and $t_m^j$ the time of event $m$ in time series $j$ with $l = 1,2,\dotsc,s_i$ and $m=1,2,\dotsc,s_j$, where $s_i$ and $s_j$ denote the number of events in the respective series.

\subsection{Event Synchronization (ES) \label{subsec:es}}

\begin{figure}
\includegraphics[width=\columnwidth]{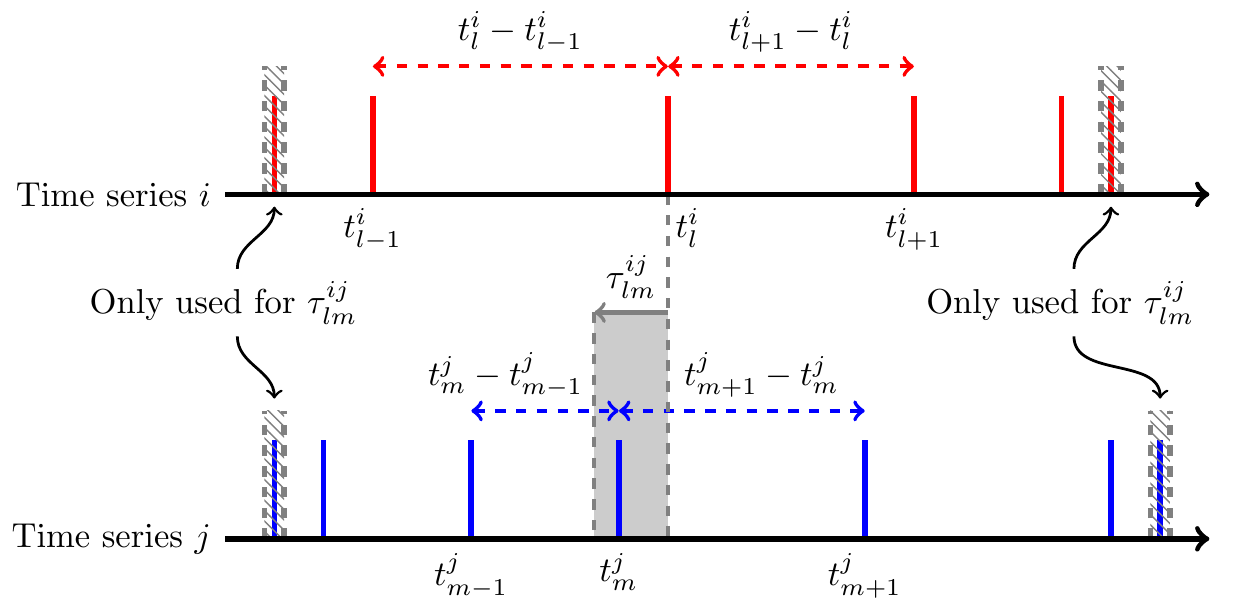}
\caption{Schematic illustration of event synchronization. \label{fig:es}}
\end{figure}

As mentioned previously, ES was first introduced by \textcite{QuianQuiroga2002} as a parameter free method for the analysis of synchronization phenomena in spiky electroencephalography (EEG) data, but has recently been applied to other fields of research as well. Fig.~\ref{fig:es} schematically illustrates the basic idea behind ES.

Two events at $t_l^i$ and $t_m^j$ are considered synchronized if they both occur within a certain data adaptive time interval of width $\tau_{lm}^{ij}$ defined as
\begin{equation} \label{eq:tau}
\tau_{lm}^{ij} = \frac{1}{2} \min \left\lbrace t_{l+1}^i - t_l^i, \, t_l^i - t_{l-1}^i, t_{m+1}^j - t_m^j, \, t_m^j - t_{m-1}^j \right\rbrace,
\end{equation}
\noindent
with $l = 2, 3, \dotsc, s_i-1$ and $m = 2, 3, \dotsc, s_j-1$ so that $\tau_{lm}^{ij}$ is not evaluated for the first and last event in order to ensure proper consideration of the boundaries.

Equation \eqref{eq:tau} implies that the more rarely events occur in either (or both) of the time series, the larger $\tau_{lm}^{ij}$ will be, so that we refer to it as a \textit{dynamic (local) coincidence interval}. Thus, if events are rare in the vicinity of either of the two events, larger deviations from an instantaneous coincidence might still be considered synchronized. The factor of $1/2$ is included to avoid double counting by making $\tau_{lm}^{ij}$ less or equal than half the minimum inter-event waiting time. The dynamic nature of $\tau_{lm}^{ij}$ simplifies the separation of independent events, which in turn results in a variety of temporal scales to be captured by a single measure. The trade-off is that, by design, the value of $\tau_{lm}^{ij}$ constantly changes between different pairs of events.

Counting the number of synchronized event occurrences in $i$ given an event in $j$ yields
\begin{equation} \label{eq:cij}
c(i|j) = \sum_{l=2}^{s_i-1} \sum_{m=2}^{s_j-1} J_{lm}^{ij},
\end{equation}
\noindent
where $J_{ij}$ is a counting function that incorporates $\tau_{lm}^{ij}$ and depends on whether or not the synchronization condition
\begin{equation} \label{eq:klmij}
\sigma_{lm}^{ij} = \left\{
\begin{array}{cll}
1 & \text{if} & 0 < t_l^i - t_m^j \leq \tau_{lm}^{ij} \\
0 & \text{else} & \\
\end{array} \right.
\end{equation}
\noindent
is met for the considered and neighbouring events:
\begin{equation} \label{eq:Jij}
J_{lm}^{ij} = \left\{
\begin{array}{cll}
1 & \text{if} & \sigma_{lm}^{ij} = 1 \\
& \text{and} & \sigma_{m,l-1}^{ji} = 0 \\
& \text{and} & \sigma_{m+1,l}^{ji} = 0 \\
1/2 & \text{if} & t_l^i = t_m^j \\
& \text{or} & \sigma_{lm}^{ij} = 1 \ \text{and} \\
& & (\sigma_{m,l-1}^{ji} = 1 \ \text{or} \\
& & \sigma_{m+1,l}^{ji} = 1) \\
0 & \text{else} &
\end{array} \right.
\end{equation}

We note that the counting function in Eq.~\eqref{eq:Jij} deviates from the original definition of ES and admittedly looks rather cumbersome. For correct specification, the changes are inevitable, though, as otherwise erroneous double counting might occur. Due to the condition of an inter-event distance that is smaller than, \textit{or equal to,} the dynamical coincidence interval $\tau_{lm}^{ij}$ in both directions, in the original definition events could be counted twice. In order to avoid this, we thus need to check for all event pairs whether one of the events has already been counted as synchronized in the opposite direction. If this is the case, a weight of 1/2 is assigned to this pair so that normalization is again carried out correctly. Such a situation can only occur if $t_l^i - t_m^j = \tau_{lm}^{ij}$ and the respective events then contribute equally to $c(i|j)$ and $c(j|i)$. While this correction should always be incorporated, it is especially important for time series with comparably low temporal resolution (like daily values of some climate variable). Alternatively, it is possible to exchange $\leq$ for $<$ in Eq.~\eqref{eq:klmij}, which has been done in later applications of the ES concept \cite{Kreuz2015,Mulansky2015}. However, this leads to an entirely new measure, called `SPIKE-Synchronization', with different aggregation and normalization. Because our focus here lies on revealing potential shortcomings that result from application of ES in its original form, we leave these more recent developments aside and restrict ourselves to a correction of the original ES measure, which has seen extensive use as mentioned earlier. Furthermore, we presume that for high resolution time series the differences will likely be small as the fundamental functioning of the dynamical coincidence interval is left unchanged.

By full analogy, we further define $c(j|i)$ and infer the \textit{strength of event synchronization} between $i$ and $j$ as
\begin{equation} \label{eq:Q}
Q_{ij}^{ES} = \frac{c(i|j) + c(j|i)}{\sqrt{(s_i-2) (s_j-2)}},
\end{equation}
\noindent
which is normalized so that $0 \leq Q_{ij}^{ES} \leq 1$, where $Q_{ij}^{ES}=1$ implies complete event synchronization and $Q_{ij}^{ES}=0$ the absence of any synchronized events. 

For the purpose of generating a functional network representation of a set of time series, we consider the pairwise ES strength as a generic statistical association measure, the estimated values of which provide the coefficients of a matrix $\mathbf{Q}^{ES} = (Q_{ij}^{ES})$. Since $Q_{ij}^{ES}$ as defined above is symmetric with respect to interchanges between $i$ and $j$, this matrix is square symmetric and can therefore be used to construct an undirected network from multivariate event data (see Section~\ref{subsec:net}).

As a simple example of the proposed modifications to the original ES definition, consider an alternating event time series, e.g., $t^i \in\{1,3,5,7,9,11\}$ and $t^j \in\{2,4,6,8,10,12\}$ with $\tau_{lm}^{ij} = 1 \, \forall \, l,m$. Using the original definition would yield $c(i|j) = 3$, $c(j|i) = 4$ and therefore $Q_{ij}^{ES} = \frac{3+4}{\sqrt{4 \cdot 4}} = 1.75$, which is not normalized as required. Our corrected version instead yields $c(i|j) = 1.5$, $c(j|i) = 2$ and thus $Q_{ij}^{ES} = 0.875$, because for every event pair at least one event is synchronized with another event in the opposite direction. Note that in perfectly alternating event time series the value $Q_{ij}^{ES}$ will always be less than 1 because the first event in one time series (here in $i$) cannot be synchronized to any event in the other time series (here $j$) in the calculation of $c(i|j)$. The ES strength between two completely synchronized event sequences, however, always yields 1, e.g., $Q_{ii}^{ES} = Q_{jj}^{ES} = 1$.


\subsection{Event Coincidence Analysis (ECA) \label{subsec:eca}}

\begin{figure}
\includegraphics[width=\columnwidth]{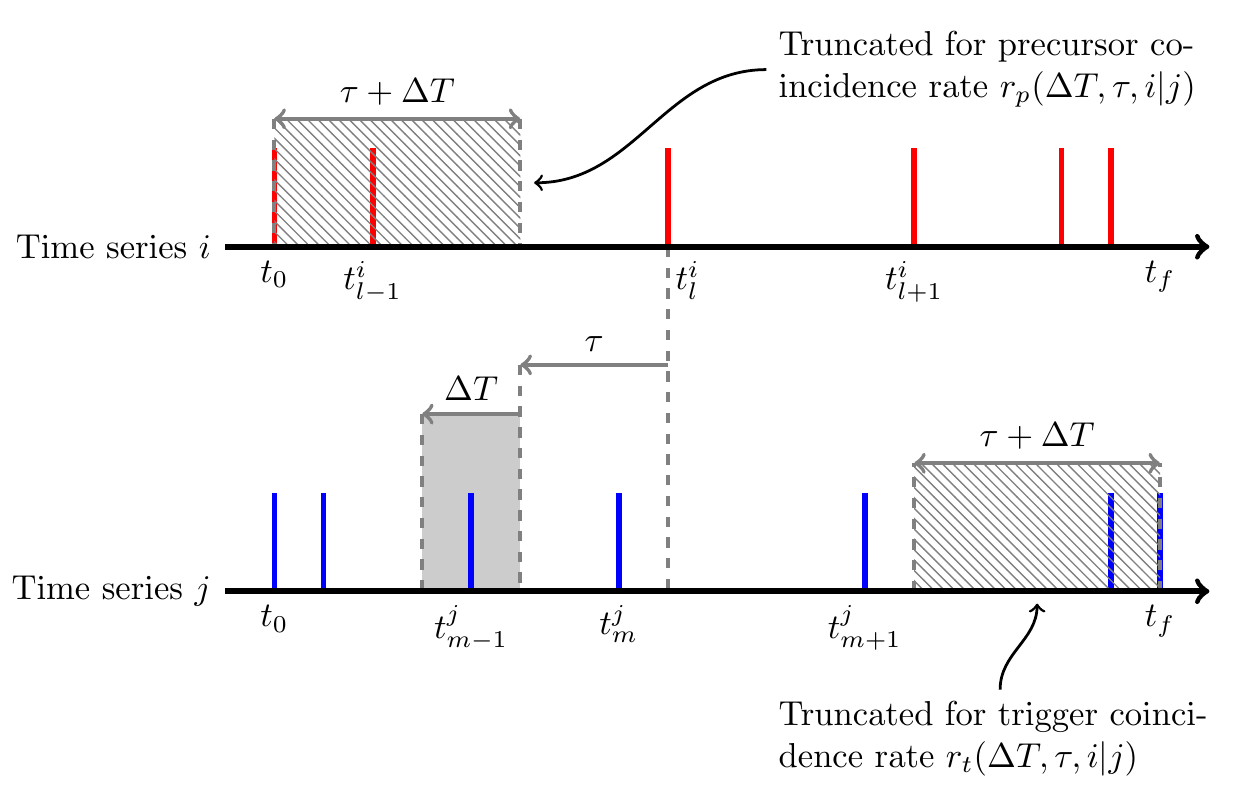}
\caption{Schematic illustration of event coincidence analysis. \label{fig:eca}}
\end{figure}

Event Coincidence Analysis (ECA) \cite{Donges2011,Donges2016} is a recently developed method to measure similarities between event time series by incorporating \emph{static (global) coincidence intervals} (as opposed to the dynamic coincidence interval of ES) and potentially also mutual time lags (which have been rarely considered along with ES in the past, but could be included here as well). Fig.~\ref{fig:eca} illustrates the general idea of ECA.

In the following, we will explain the estimation of event coincidence rates in full detail only for the case where events in $j$ precede events in $i$. Event coincidence rates for the other direction are obtained by simply exchanging $i$ and $j$ in all formulas. Again, let $t_l^i$ denote the time index of event $l$ in series $i$ and $t_m^j$ the time index of event $m$ in series $j$, but now again with $l = 1, 2, \dotsc, s_i$ and $m = 1, 2, \dotsc, s_j$. All events have been observed during the same observation period $[t_0,t_f]$. We define an \textit{instantaneous event coincidence} if two events at $t_l^i$ and $t_m^j$ occur within a certain static coincidence interval $\Delta T$ so that $0 \leq t_l^i - t_m^j \leq \Delta T$. Accordingly, a \textit{lagged event coincidence} is a coincidence between a time shifted event at $t_l^i - \tau$ and an event at $t_m^j \leq t_l^i - \tau$, with a given \textit{time lag} $\tau \geq 0$, implying that $0 \leq (t_l^i - \tau) - t_m^j \leq \Delta T$ holds.

To quantify the degree of synchrony between event in the two time series $i$ and $j$, we distinguish between so-called \textit{precursor} and \textit{trigger} event coincidence rates \cite{Schleussner2016,Donges2016}. This distinction relates to the question whether the number of events in $i$ or $j$ are used for normalization: the precursor rate quantifies the fraction of events in $i$ that have been preceded by at least one event in $j$, while the trigger rate gives the fraction of events in $j$ than have been followed by at least one event in $i$. 

Formally, the \textit{precursor event coincidence rate} is thus defined as
\begin{eqnarray} \label{eq:precursor}
r_p ( && i|j; \Delta T, \tau) = \nonumber \\*
&& \frac{1}{s_i - s_i'} \sum_{l=1+s_i'}^{s_i} \Theta \left\lbrace \sum_{m=1}^{s_j} 1_{[0,\Delta T]} \left( (t_l^i - \tau) - t_m^j \right) \right\rbrace.
\end{eqnarray}
Since it measures the fraction of ($\tau$-lagged) events in $i$ that are preceded by \textit{at least} one event in $j$, multiple events in $j$ falling within the static coincidence interval $\Delta T$ relative to an event in $i$ are only counted once. Here, $\Theta(\bullet)$ denotes the left-continuous Heaviside step function with $\Theta(0)=0$, which rules out double counting. $1_I(\bullet)$ represents the indicator function of the interval $I$ defined as
\begin{equation}
1_I(x) = \left\{
\begin{array}{cll}
1 & \text{if} & x \in I \\
0 & \text{if} & x \notin I \\
\end{array} \right. .
\end{equation}
\noindent
Here, the integer $s_i'$ denotes the number of events in $i$ that occur within the interval $[t_0, t_0 + \tau + \Delta T]$, i.e., 
\begin{equation}
s_i' = \sum_{l=1}^{s_i} 1_{[t_0,t_0+\tau+\Delta T]} (t_l^i).
\end{equation}
\noindent
By full analogy, we define $r_p(j|i; \Delta T, \tau)$ as the precursor event coincidence rate for the case of events in $j$ being preceded by at least one event in $i$. 

In turn, by taking the events in $j$ as the basis for normalization, we obtain the \textit{trigger event coincidence rate}
\begin{eqnarray} \label{eq:trigger}
r_t ( && i|j; \Delta T, \tau) = \nonumber \\*
&& \frac{1}{s_j-s_j''} \sum_{m=1}^{s_j-s_j''} \Theta \left\lbrace \sum_{l=1}^{s_i} 1_{[0,\Delta T]} \left( (t_l^i - \tau) - t_m^j \right) \right\rbrace,
\end{eqnarray}
\noindent
which measures the fraction of events in $j$ that are followed by at least one event in $i$. Here, $s_j''$ counts the number of events in $j$ that occur within the interval $[t_f - (\tau + \Delta T), t_f]$:
\begin{equation}
s_j'' = \sum_{m=1}^{s_j} 1_{[t_f - (\tau + \Delta T), t_f]} (t_m^j)
\end{equation}
\noindent
The primes in $s_i'$ and $s_j''$ are intended to avoid confusion once the indices are swapped to calculate the same types of coincidence rates in opposite directions. One prime refers to the number of events in the interval at the beginning of the time series, while two primes refer to the number of events in the interval at the end. Finally, we can define $r_t(j|i; \Delta T, \tau)$ as the trigger coincidence rate for the opposite case of events in $i$ preceding events in $j$. Clearly, being defined as fractions, all four possible event coincidence rates are confined to values between 0 and 1.

The idea to truncate the succeeding time series at the beginning for the precursor event coincidence rate and the preceding time series at the end for the trigger event coincidence rate provides a necessary correction of the original ECA definition \cite{Donges2016}. Ignoring this may again lead to an erroneous normalization especially if $\tau \neq 0$, because for the precursor event coincidence rate, events in $i$ before $t_0 + \tau$ can never coincide with any event in $j$. Similarly, for the trigger event coincidence rate, events in $j$ after $t_f - \tau$ can never coincide with any event in $i$. If this is disregarded, the respective event coincidence rate may end up at a value below 1, even if all events that could possibly coincide also do coincide. This might lead to an underestimation, but at least not to values larger than 1 as for the uncorrected ES definition. The committed error without this correction vanishes with long time series as the number of events becomes sufficiently large. However, for finite event sequences, proper truncation should be employed.

Altogether, ECA yields four event coincidence rates for every pair of event time series, namely the precursor and trigger event coincidence rates in both directions. In what follows, we will only use the trigger event coincidence rates as the differences with respect to the precursor event coincidence rates have been found to be commonly very small across all numerical investigations presented in the following. However, it should not be argued that this is necessarily always the case by construction. Consider, as a simple thought example, two event sequences with the same number of events. In the first series, all events occur with a precise clock after a fixed waiting time, and every second of those events is followed by two events in the other series that occur in close succession (within $\Delta T$). In this case, each event of the second series has a precursor in the first series, while only every second event of the first series triggers events in the second one.

In order to obtain a single statistical association measure $Q_{ij}^{ECA}$ as the \textit{degree of event synchrony}, we can employ either the mean or the maximum value of the two directed trigger event coincidence rates $r_t(i|j; \Delta T, \tau)$ and $r_t(j|i; \Delta T, \tau)$, denoted as $Q_{ij}^{ECA,mean}$ or $Q_{ij}^{ECA,max}$, respectively. The former is especially appropriate for bidirectional dependencies, whereas the latter emphasizes the strengths of unidirectional dependencies more strongly, irrespective of the direction.

Finally, similar to the ES, we can create a similarity matrix from pairwise values of event coincidence rates if more than two simultaneously measured time series are available. The resulting similarity matrix $\mathbf{Q}^{ECA} = (Q_{ij}^{ECA})$ of either the mean or maximum values is again normalized to $0 \leq Q_{ij}^{ECA} \leq 1$ and square symmetric.

Note that for simplicity and tractability of our main argument, we herein refrain from exploring information on directionalities in both ES and ECA. In the context of ES, additional directionality information can be obtained by considering the differences between $c(i|j)$ and $c(j|i)$ \cite{QuianQuiroga2002,Malik2010,Malik2012}. In turn, for ECA either a similar difference between any of the two event coincidence rates under an exchange of $i$ and $j$ or the distinction between trigger and precursor event coincidence rates could be used for a similar purpose. We outline corresponding further theoretical studies as a subject of future work.


\section{Numerical study: Bivariate AR(1) processes \label{sec:artificial}}

We first compare ES and ECA for artificial data stemming from a simple bivariate first order vector autoregressive (VAR(1)) process. 
A comprehensive, yet not completely exhaustive, comparison of different synchronization measures for dynamical systems has been provided in \textcite{Kreuz2007a}, but only included ES and not ECA, and only compared the different approaches with regard to coupling and noise strength, but not to serial dependency and particularly event clustering, which remains our focus in the following.

\begin{figure}
\includegraphics[width=\columnwidth]{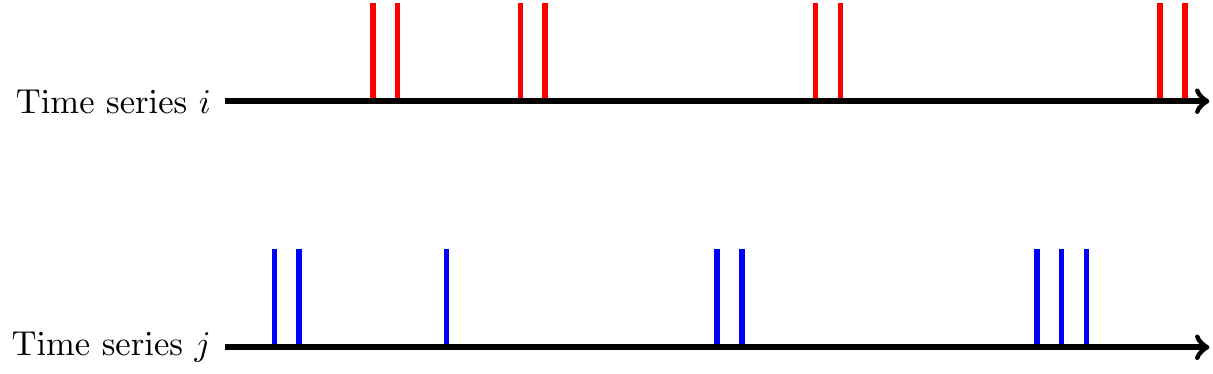}
\caption{Schematic illustration of an example of clustered event time series. \label{fig:prob}}
\end{figure}

Fig.~\ref{fig:prob} shows a stylized example where events tend to occur in pairs at subsequent time steps. Evidently, there is some form of lagged synchronization since a sequence of one to three consecutive events in $j$ is always followed by an event pair in $i$. However, ES is unable to detect this type of synchronization, because the dynamical coincidence interval $\tau_{lm}^{ij}$ always remains small due to the short inter-event distances. The illustrative result of $Q_{ij}^{ES}=0$ is unsatisfactory and highlights a major caveat of ES: it fails to properly unravel different temporal scales if events are clustered in time. By contrast, given appropriate parameter values for $\Delta T$ and $\tau$, ECA would clearly detect a directional relationship so that $Q_{ij}^{ECA} = 1$.

Although, for a juxtaposition of two nonlinear methods, it might appear sensible to analyze coupled nonlinear systems, a simple linear VAR(1) process already suffices to demonstrate the fundamental differences between ES and ECA with regard to event clustering. This approach is also much easier to comprehend, whereas nonlinear systems often defy straightforward imagination. Therefore, we use a VAR(1) model
\begin{align}
x_t &= \varphi_1 x_{t-1} + \kappa_1 y_{t-1} + \varepsilon_{1,t} \label{eq:var1} \\
y_t &= \varphi_2 y_{t-1} + \kappa_2 x_{t-1} + \varepsilon_{2,t}
\end{align}
\noindent
to create artificial time series that depend on autoregressive parameters $\varphi_1$ and $\varphi_2$ (modeling serial correlations) and coupling parameters $\kappa_1$ and $\kappa_2$ (modeling cross correlations) for two variables $x_t$ and $y_t$. The error terms $\varepsilon_{1,t}$ and $\varepsilon_{2,t}$ follow two independent standard normal distributions with mean $\mu=0$ and variance $\sigma=1$, i.e., $\varepsilon_{1,t} \sim \mathcal{N}(0,\,1)$ and $\varepsilon_{2,t} \sim \mathcal{N}(0,\,1)$. Similarly, the initial values $x_1$ and $y_1$ are also drawn from two independent standard normal distributions.

For a given realization of this bivariate stochastic process, the time steps where the associated values exceed a given percentile threshold yield two event time series, which can be used as an input for both ES and ECA. As an illustrative example, for ECA we set the parameters to $\tau=0$ and $\Delta T=3$. We simulate with a time series length of 1,000 points and threshold at the 90th percentile, yielding $s_i = s_j = 100$ events per time series. We consider 1,000 independent runs and calculate the averages of $Q_{ij}^{ES}$, $Q_{ij}^{ECA,mean}$ and $Q_{ij}^{ECA,max}$ over $i,j = 1,\dotsc,1,000$, respectively. These ensemble averages are denoted as $\overline{Q}^{ES}$, $\overline{Q}^{ECA,mean}$ and $\overline{Q}^{ECA,max}$, respectively, all of which are functions of $\varphi_1$, $\varphi_2$, $\kappa_1$ and $\kappa_2$. Following our previous considerations (see Fig.~\ref{fig:prob}), we expect that ES will show an undesirable negative dependency on $\varphi_1$ and $\varphi_2$. However, simulating the time series for a discrete set of varying parameters with $\varphi_1, \varphi_2, \kappa_1, \kappa_2 \in \left\lbrace 0,0.05,0.1,0.15,\dotsc, 0.95, 1 \right\rbrace$ together with the choice of using either $Q_{ij}^{ECA,mean}$ or $Q_{ij}^{ECA,max}$ entails (too) many degrees of freedom for a meaningful analysis. In order to reduce computational efforts and focus on the most relevant aspects, we consider here only two illustrative settings, one simplified bivariate and one extreme univariate case.

\begin{figure*}
    \includegraphics{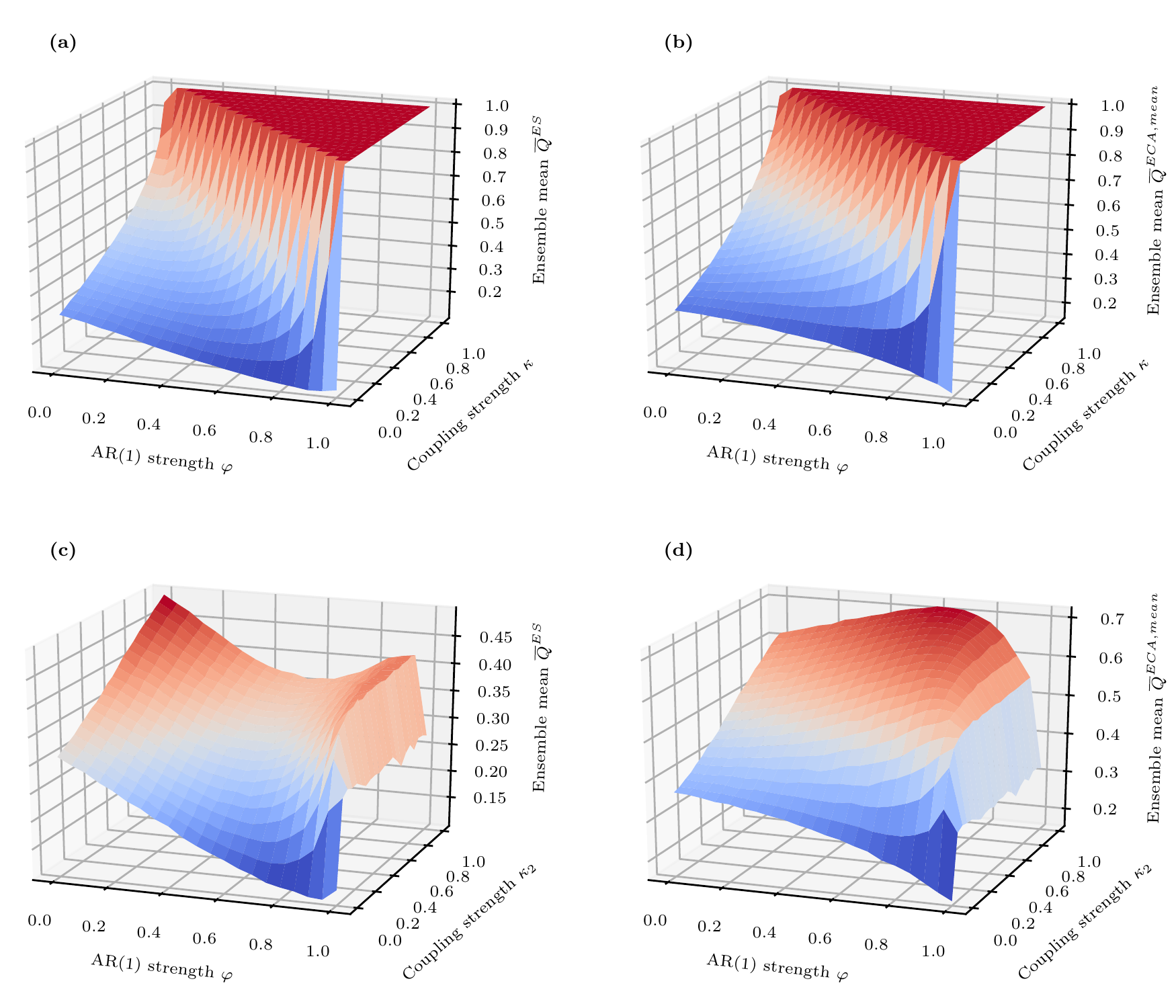}
    \caption{Ensemble means of the statistical association measures provided by ES and ECA for simplified bivariate and univariate cases of the bivariate VAR(1) model. \textbf{(a)}: Bidirectional coupling, $\varphi:=\varphi_1=\varphi_2$, $\kappa:=\kappa_1=\kappa_2$, ES. \textbf{(b)}: Bidirectional coupling, ECA. \textbf{(c)}: Unidirectional coupling, $\varphi:=\varphi_1=\varphi_2$, $\kappa_1=0$, ES. \textbf{(d)}: Unidirectional coupling, ECA. \label{fig:es_eca_bidir_unidir}}
\end{figure*}

Fig.~\ref{fig:es_eca_bidir_unidir} shows four 3D surface plots of the ensemble means $\overline{Q}^{ES}$ and $\overline{Q}^{ECA,mean}$ in dependence of the AR(1) and coupling parameters. The top row with subplots (a) and (b) contains the simplified bidirectional case, for which we set $\varphi:=\varphi_1=\varphi_2$ and $\kappa:=\kappa_1=\kappa_2$. The bottom row with subplots (c) and (d) contains the extreme unidirectional case, for which we again set $\varphi:=\varphi_1=\varphi_2$, but now $\kappa_1=0$ so that $\kappa_2=\kappa$ remains the only free coupling parameter. For brevity, we leave out the results of $\overline{Q}^{ECA,max}$ as they turned out to be qualitatively indistinguishable from $\overline{Q}^{ECA,mean}$. Note that the plots do not share the same $z$ axes as the absolute values are not directly comparable among the different measures.

\begin{figure}
    \includegraphics{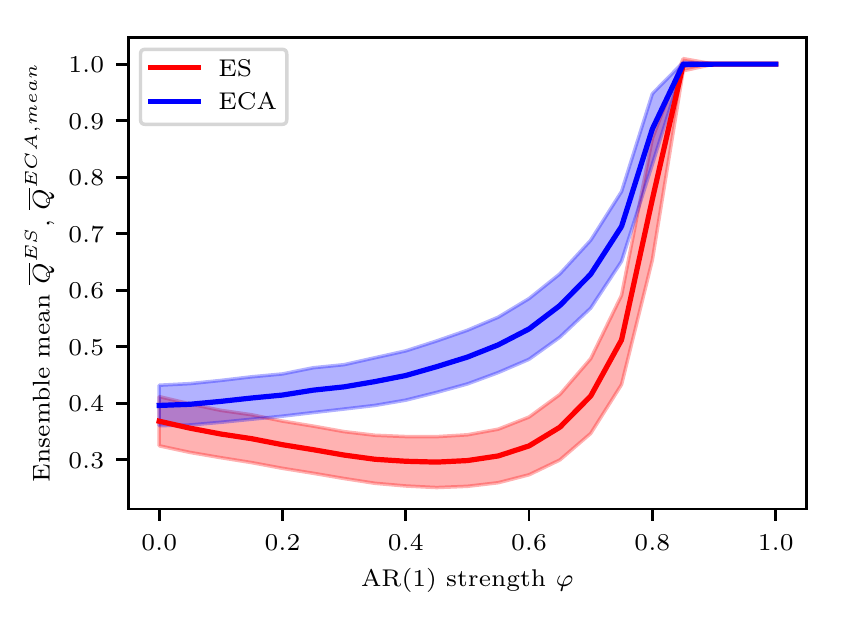}
    \caption{Transect of the simulation results for the symmetrically bidirectionally coupled AR(1) processes (Fig.~\ref{fig:es_eca_bidir_unidir}a,b) with $\kappa=0.2$. Solid lines and shadings indicate the ensemble means and standard deviations, respectively. \label{fig:es_eca_bidir_trans}}
\end{figure}

Looking at the simplified bidirectional case (Fig.~\ref{fig:es_eca_bidir_unidir}a,b), we see that for $\varphi + \kappa \geq 1$ the results of both methods equal 1, since the VAR(1) process becomes nonstationary so that $x_t$ and $y_t$ diverge and all events occur at subsequent time steps. This leads to a perfect, albeit meaningless, synchronization. Much more interesting is the behavior of ES and ECA in dependence on the AR(1) parameter $\varphi$, which controls the serial dependency and, hence, temporal clustering of events. For any $\kappa \neq 0$, we observe that $\overline{Q}^{ES}$ first decreases with $\varphi$, whereas $\overline{Q}^{ECA,mean}$ continuously increases with $\varphi$. This is depicted more clearly in Fig.~\ref{fig:es_eca_bidir_trans}, where a transect of both panels at $\kappa=0.2$ is shown together with the respective ensemble standard deviations. The results confirm our initial expectation that ES is adversely affected by serial dependencies, which are here parameterized via the AR(1) parameter $\varphi$ as justified below. On the other hand, ECA values show a strictly positive dependence on $\varphi$. This is a much more understandable and meaningful behaviour as increasing $\varphi$ also increases statistical persistence, which makes events occur less erratic, but rather in temporal clusters. While this does not justify a stronger synchronization as such, it does lead to the fact that if a synchronization occurs it is more likely to include several events at once. As normalization in ECA is carried out over the number of individual events in the time series, increasing serial dependencies through $\varphi$ ultimately increases the event coincidence rate (as it is also common in other classical statistical association measures like Pearson correlation). Such reasoning would also be plausible for ES, but clearly does not become apparent from our numerical results.

Turning to the extreme unidirectional case (Fig.~\ref{fig:es_eca_bidir_unidir}c,d), the strong bias of ES in the presence of serial dependencies stands out even more prominently as $\overline{Q}^{ES}$ decreases with respect to $\varphi$ for all values of $\kappa_2$ up to $\varphi \approx 0.7$. Quite contrarily, the values of $\overline{Q}^{ECA,mean}$ increase slowly but steadily for almost all values of $\kappa_2$ up to $\varphi \approx 0.7$. These patterns match the expected behavior, underpinning our conceptual concerns regarding the potential caveats of ES and providing indications in support of ECA as a more reliable measure of event synchronicity. Yet, for $\varphi \gtrsim 0.7$, the results show intriguing patterns. The ES values now increase again with rising $\varphi$, only to drop abruptly at $\varphi=1$. The initial increase is likely a result of increasing persistence that eventually leads to the expected pattern that is commonly observed for ECA, but only commences for very high $\varphi$ in ES. The abrupt drop appears to be a statistical artifact stemming from the nonstationarity of the process at $\varphi=1$. In turn, for ECA, the obtained values above $\varphi \gtrsim 0.7$ reverse the initial pattern by falling again. As statistical persistence increases with $\varphi$, events in both $x_t$ and $y_t$ form more and more clusters in a certain part of the underlying time series. For $x_t$, which is entirely independent of $y_t$ as $\kappa_1=0$, this means that the probability of event clusters falling into the same period as in $y_t$ accordingly declines. For $y_t$, even a large unidirectional coupling parameter $\kappa_2$ does not curtail this outcome substantially so that the overall mean $\overline{Q}^{ECA,mean}$ also declines. Altogether, the region above $\varphi \approx 0.7$ in a completely unidirectional setting is a particularly extreme regime that is intellectually interesting to scrutinize, but not crucial for the overall interpretation of evolving tendencies.

\begin{figure}
    \includegraphics{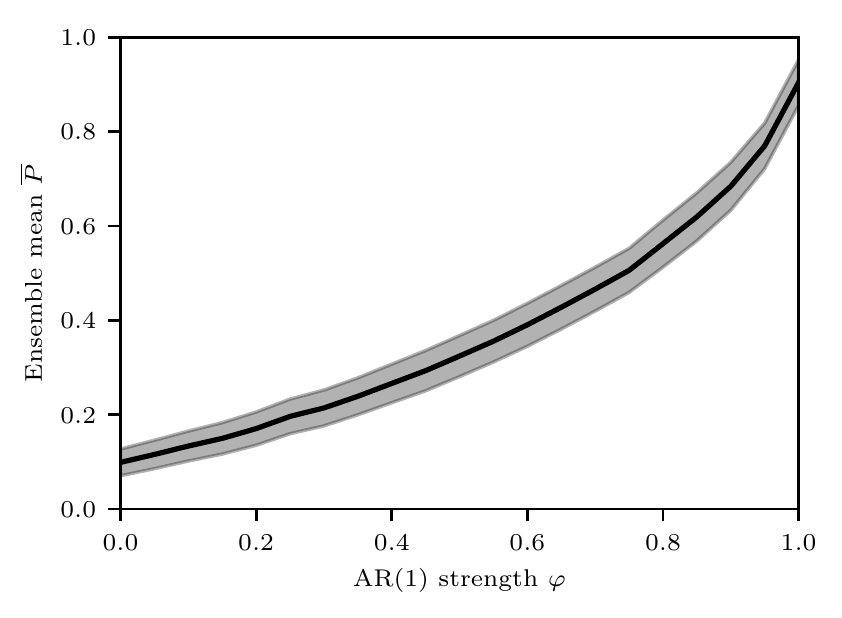}
    \caption{Dependency of pairing coefficient $P_i$ on AR(1) strength $\varphi$, shown as ensemble mean (solid line) and standard deviation (shaded band).
    \label{fig:phi_pc}}
\end{figure}

In order to further confirm the effect of $\varphi$ stipulated above on the inter-event distances that ultimately lead to the described behavior in both considered cases, we define a simple measure for event clustering, the \textit{pairing coefficient} $P_i$, as the fraction of events occurring on subsequent time steps,
\begin{equation} \label{eq:pair}
P_i = \frac{1}{s_i-1} \sum_{l=1}^{s_i-1} \delta \left[ \left( t_{l+1}^i - t_l^i \right) - 1 \right],
\end{equation}
\noindent
so that $0 \leq P_i \leq 1$, where $P_i=0$ means that no event pairs form at all and $P_i=1$ that all events occur on consecutive time steps. Note that $P_i$ is a univariate measure characterizing the properties of a single time series $i$ that essentially entails the first value ($\Delta t=1$) of the respective empirical inter-event distance distribution. Thus, $P_i$ allows us to scrutinize the characteristics of the input data used for estimating ES and ECA in a simple manner without any notion of coupling or synchronization. Fig.~\ref{fig:phi_pc} shows the dependency of the ensemble mean of $P_i$, denoted as $\overline{P}$, on the AR(1) strength $\varphi$, with calculations based on 1,000 independent realizations as before. Clearly, $\overline{P}$ is strictly monotonically increasing with $\varphi$. At high values of $\varphi$, events exhibit strong serial dependencies, thereby increasing statistical persistence. This confirms our reasoning of using $\varphi$ as a proxy for serial dependency in the resulting event time series by providing the missing, but expected, link. The pairing coefficient will be drawn upon later again.

Even though the absolute difference between ES and ECA results might not appear overwhelming for the bivariate case, it should be noted that the respective quantities should only be compared to each other in relative terms as they are usually ranked internally, before being referred to. In functional network analysis, for instance, the values of ES and ECA would be ranked so that only the strongest links are included in the network representation \cite{Donner2017}. This implies that even small changes in the corresponding association values might entail large consequences for the inferred network structure. Thus, it is not the absolute values that should be compared, but rather the patterns in response to changes in $\varphi$ and $\kappa$ as just described.

Taken together, our simulation results confirm our initial expectation that the dynamical coincidence interval $\tau_{lm}^{ij}$ unambiguously renders ES insensitive to properly detect synchronization if the events in a time series are strongly clustered, which is a common property of climate extremes \cite{Kropp2011}. This undesirable outcome may hamper the reliability of results and interpretations obtained from such networks, as we will show in the following.

 \section{Real world examples \label{sec:realworld}}

Following our numerical results for the artificial data in Section~\ref{sec:artificial}, we next attempt to shed some more light on the real world implications of those findings. Considering the extensive previous research on functional network analysis based on ES as a statistical similarity measure, Section \ref{subsec:net} concisely reviews the key elements of network construction and analysis. Then, in Section \ref{subsec:climdata}, we demonstrate that ES yields biased results for climate network representations of heavy rainfall events, since climate time series commonly exhibit serial dependencies and clustering among extreme events. Along with reproducing some results of previous studies based on ES, we present substantially different results based on ECA that are not only more robust in the presence of event clustering, but also allow us to analyze the temporal evolution of extreme events in a functional precipitation network, enabling worthwhile customized analyses on a more detailed level than when using ES.

As an illustrative counterexample, Section \ref{subsec:eeg} analyzes five sets of bivariate rat EEG signals, including one non-epileptic example and four epileptic spike trains, for which ES and ECA yield qualitatively very similar results. This highlights that due to a relatively narrow waiting time distribution of clearly discernible EEG spikes (i.e., the existence of a relatively regular internal pacemaker), temporal clustering among events is not a major issue here (i.e., there exist serial dependencies among events, but of an entirely different form than in the previous numerical example). Furthermore, we compare our results in view of a different event definition that is arguably more directly applicable to the detection of local spikes in noisy EEG signals.

\subsection{Functional networks \label{subsec:net}}

\begin{figure}
    \includegraphics[width=\columnwidth]{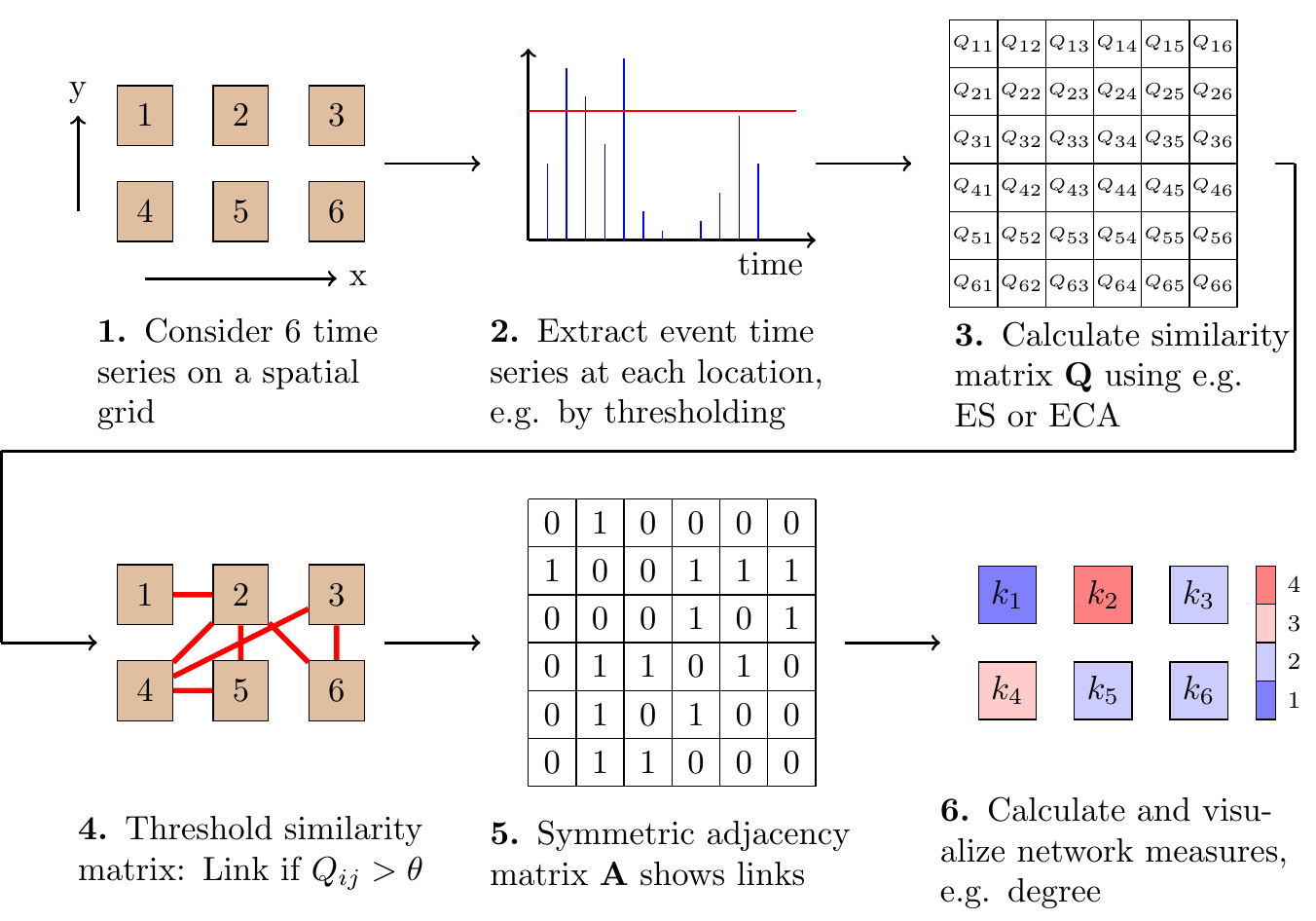}
    \caption{Flow chart of functional network analysis for a generic example of spatially embedded time series. \label{fig:climnet}}
\end{figure}

The combination of nonlinear time series analysis with complex network theory is a rapidly growing area of research as it allows to disentangle and visualize spatiotemporal patterns from large scale datasets. Fig.~\ref{fig:climnet} shows a flow chart of how to incorporate ES and ECA into the construction of a spatially embedded functional network. While this bears the hallmarks of climate network analysis \cite{Donner2017,Dijkstra2019}, it is straightforward to extend this approach to other applications.

A network $G=(V,E)$ is defined by a set of \textit{vertices} (or nodes) $V=\{1,\dotsc,N\}$ with $N=|V|$ and a set of \textit{edges} (or links) $E \subseteq V \times V$. The edges $E$ with $K=|E|$ are represented in the adjacency matrix $\mathbf{A}$, in which self connections conventionally do not exist so that $A_{ii}=0 \, \forall \, i$. Having calculated the similarity matrix $\mathbf{Q}$ from either ES or ECA, we link $i$ and $j$ if $Q_{ij}$ is above a certain threshold $\theta$. Thus, we obtain a binary square symmetric \textit{adjacency matrix} $\mathbf{A}$ of an undirected network, where $A_{ij}=A_{ji}=1$ indicates a link between $i$ and $j$ and $A_{ij}=A_{ji}=0$ the lack thereof. As a valuable alternative to choosing a particular value of $\theta$, it is common practice to instead predefine a \textit{global edge density} $\rho$, which extracts the strongest statistical associations and thereby indirectly defines $\theta$.

Subsequently, the resulting adjacency matrix can be analyzed by means of complex network theory. Out of the abundance of existing network measures \cite{Costa2007}, for the sake of brevity, in this work we only consider the \textit{degree density}
\begin{equation} \label{eq:rhoi}
\rho_i = \frac{1}{N-1} \sum_{j=1}^N A_{ij}
\end{equation}
\noindent
as the simplest and most prominent vertex measure, yielding the number of connections associated with each node, normalized to $0 \leq \rho_i \leq 1$.

\subsection{Precipitation data \label{subsec:climdata}}

Following upon previous works on ES based functional climate networks for the Indian subcontinent \cite{Malik2010,Malik2012,Stolbova2014}, we use gauge adjusted satellite data from the Tropical Rainfall Measuring Mission (TRMM)\cite{GESDISC2016} to construct a climate network representation of extreme precipitation during the Indian Summer Monsoon (ISM). We select resampled daily precipitation sums on a square grid with a spatial resolution of $0.25^{\circ} \times 0.25^{\circ}$ ($\sim \SI{28}{km}$) for the period 1998--2015 (TMPA 3B42 V7), from which we extract the summer monsoon season of June to September. For the chosen area from $60.125-99.875^{\circ}$E and $5.125-39.875^{\circ}$N, we thus have 22,400 ($160 \times 140$) grid points, each constituting a time series of 2,196 ($18 \cdot 122$) days, i.e., a total of 49,190,400 observations. In accordance with \textcite{Stolbova2014} we also threshold at the 90th percentile and select a global edge density of $\rho=0.05$ as a convenient value \cite{Donner2017}.

\begin{figure}
    \includegraphics{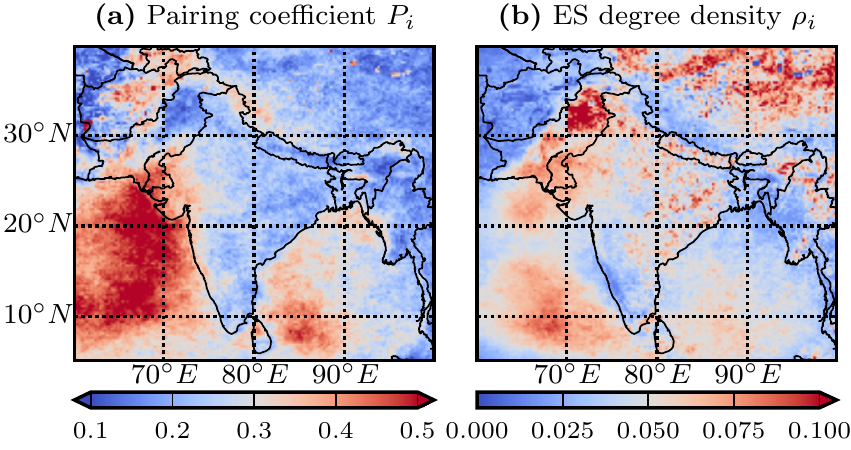}
    \caption{Spatial pattern of the pairing coefficient and the degree density of the ES based functional climate network. \label{fig:es_deg_pair}}
\end{figure}

Although the capability of ES to dynamically incorporate different time scales at once through $\tau_{lm}^{ij}$ appears worthwhile at first, it entails a major disadvantage by overemphasizing the node degree of the resulting functional climate network in regions where events are temporally isolated. This is a consequence of the systematic underestimation of the degree values in regions where events tend to occur temporally clustered and, hence, an immediate manifestation of the adverse effect of temporal event clustering on the results of ES as demonstrated in Section~\ref{sec:artificial}. In order to quantify and visualize this undesirable property, we calculate the previously introduced pairing coefficient $P_i$ for all grid points and plot this alongside the ES degree density $\rho_i$ in Fig.~\ref{fig:es_deg_pair}.

\begin{figure}
    \includegraphics{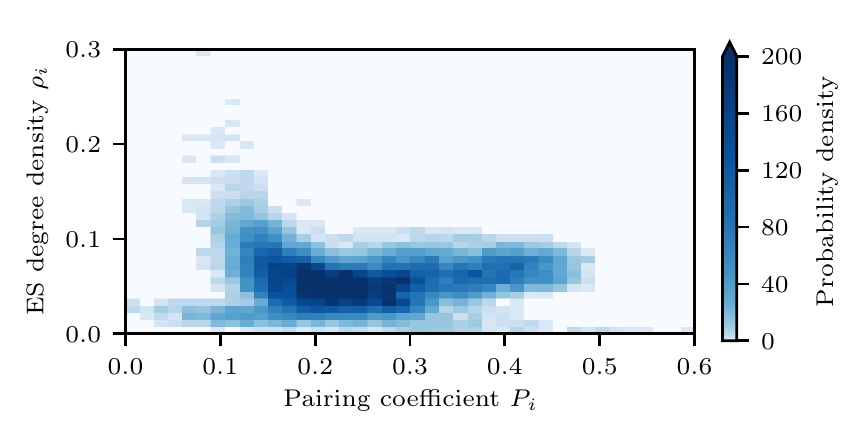}
    \caption{Estimate of the joint probability density of the pairing coefficient and the degree density of the ES based functional climate network. \label{fig:es_heat}}
\end{figure}

Notably, Fig.~\ref{fig:es_deg_pair}b reproduces the results of \textcite[][Fig.~3]{Stolbova2014} almost exactly, with minor differences originating from the described corrections of the ES algorithm and a slightly longer time period covered. Furthermore, our results are also very similar to those of \textcite{Malik2010, Malik2012}, who used a different data source without ocean coverage. 

Although Fig.~\ref{fig:es_deg_pair}a solely contains local information of the dynamical characteristics of events at each individual grid point without any notion of coupling to other grid points whatsoever, we observe interesting similarities in comparison with Fig.~\ref{fig:es_deg_pair}b. Specifically, in areas where the degree density is high, the pairing coefficient very often has low values, and vice versa. In many areas, the two figures resemble negative images of each other. This holds especially true for regions that had been reported in previous studies as important for governing monsoon dynamics at large spatial scales, such as Northern Pakistan, the Tibetan Plateau, or the Eastern Ghats \cite{Malik2012, Stolbova2014}. Note that even minor artifacts such as interspersed grid points with low $\rho_i$ values on the Tibetan Plateau can exhibit a high pairing coefficient $P_i$. The effect of $P_i$ on $\rho_i$ is further displayed in a two-dimensional density plot, see Fig.~\ref{fig:es_heat}, which reveals a negative relationship between $\rho_i$ and pairing coefficient up to $P_i \approx 0.2$. Fundamentally, this means that a trivial property of the local time series often predetermines the event synchronization strength to other grid points in many areas, especially those deemed pivotal for monsoon dynamics. This suggests that ES may not be an ideal similarity measure for time series of extreme climate events, which frequently entail serial dependencies and temporal event clustering \cite{Kropp2011}. Furthermore, our observations are compatible with the artificial data results from Section~\ref{sec:artificial} and highlight the practical implications of the described weaknesses of ES.

\begin{figure}
    \includegraphics{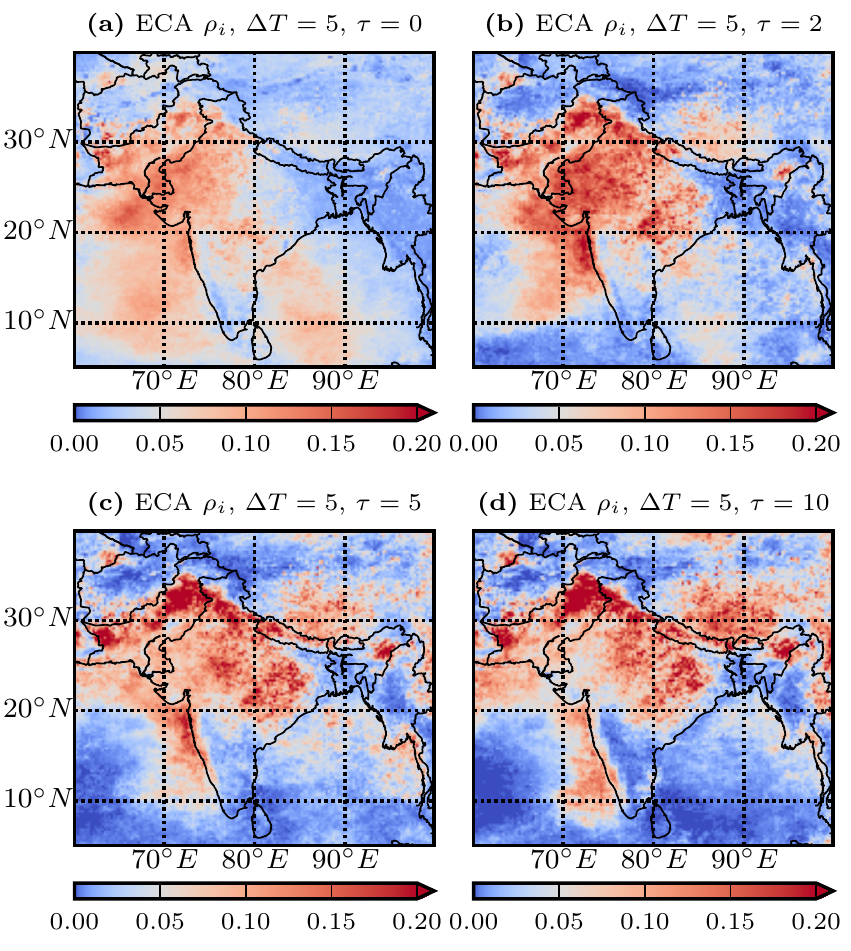}
    \caption{Spatial pattern of the degree density of the ECA based functional climate network for four different parameter settings. \label{fig:eca_deg}}
\end{figure}

\begin{figure*}
    \includegraphics{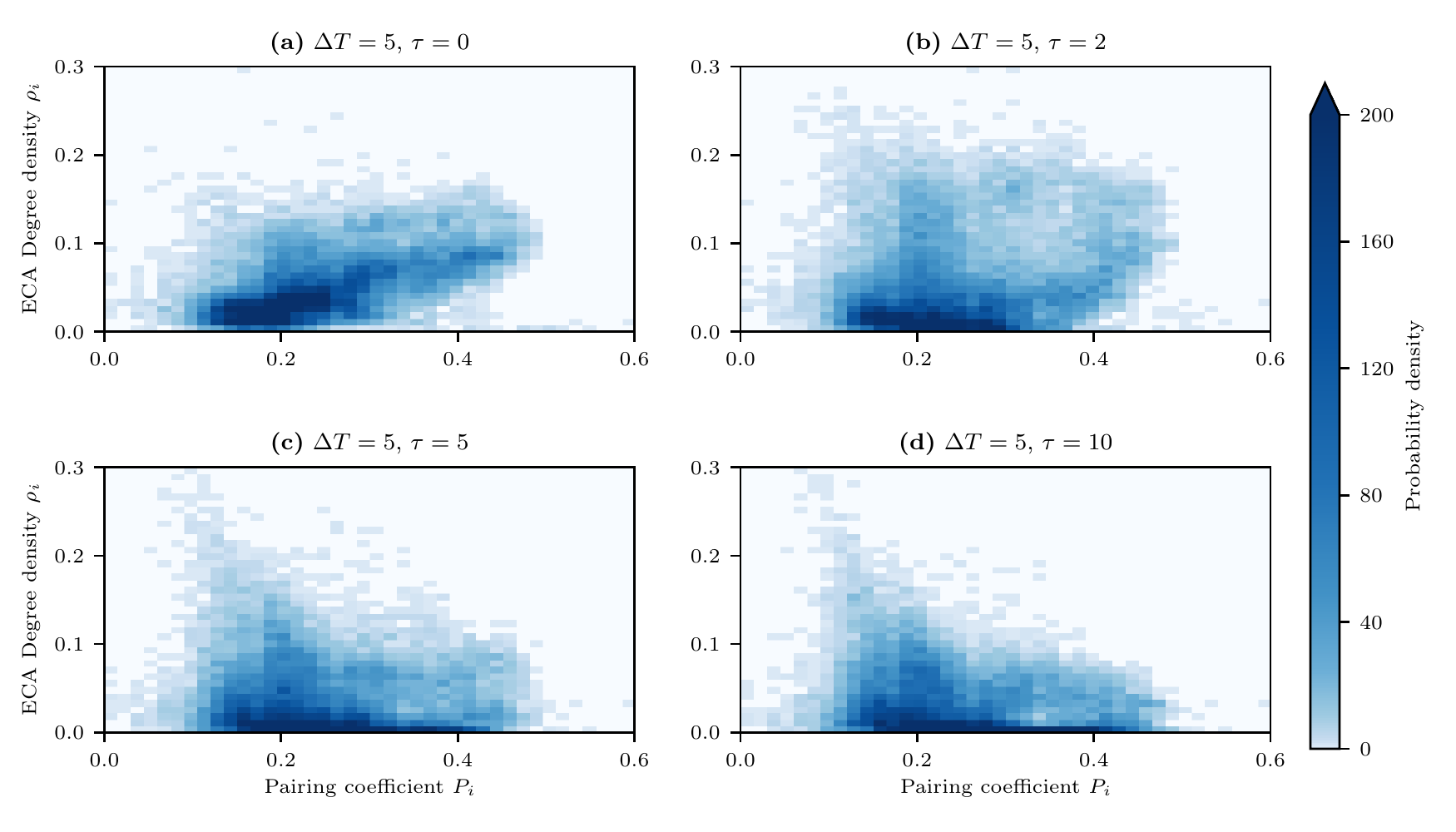}
    \caption{Estimate of the joint probability density of the pairing coefficient and the degree density of the ECA based functional climate networks. \label{fig:eca_heat}}
\end{figure*}

On the other hand, the ECA based networks do not exhibit such adverse dependencies of the node degree on the pairing coefficient. Fig.~\ref{fig:eca_deg} shows the degree density field for four different parameter settings with $\Delta T = 5$ and $\tau \in \{0,2,5,10\}$ days. Varying $\tau$ while fixing $\Delta T$ enables us to clearly isolate lagged synchrony and thus to analyze the spatiotemporal evolution of synchronous extreme precipitation events. Effectively, this moves the fixed-length synchrony detection window back in time (see Fig.~\ref{fig:eca}). This stands in contrast to fixing $\tau$ while varying $\Delta T$, which would not have allowed us to clearly disentangle lagged from instantaneous synchrony (within $\Delta T$). The obtained spatial patterns vastly differ from those in Fig.~\ref{fig:es_deg_pair}b, and the additional parameters of ECA even allow to isolate distinct time scales, thereby enabling us to assess the temporal evolution of heavy precipitation patterns along the sequence of climate networks. In our opinion, this provides a valuable extension of the ES approach, where the specific underlying time scales remain unknown, rendering the outcome of functional climate network analysis comparatively opaque. Similarly to the ES case, Fig.~\ref{fig:eca_heat} shows the influence of $P_i$ on $\rho_i$ for the four considered parameter settings of the ECA based network. In contrast to Fig.~\ref{fig:es_heat}, the relationship between both characteristics is far less pronounced as the points scatter much more. The slightly positive dependency in Fig.~\ref{fig:eca_heat}a gradually evolves into a slightly negative one in Fig.~\ref{fig:eca_heat}d, but is subject to strong dispersion.

Based on our findings as reported above, we tentatively propose ECA as the preferable similarity measure for event based functional climate network construction and analysis and stress that previous results based on ES should be interpreted with caution and carefully re-examined wherever possible. Apart from enabling a proper disentanglement of synchrony from serial dependency, we further advocate ECA's ability to precisely analyze the temporal scales encoded in a given network. Yet, even if such a refined analysis is not desired, the adverse effects of event clustering on ES can still overshadow the potential benefits of incorporating multiple time scales at once to a large extent.

\subsection{EEG data \label{subsec:eeg}}

\begin{figure*}
    \includegraphics{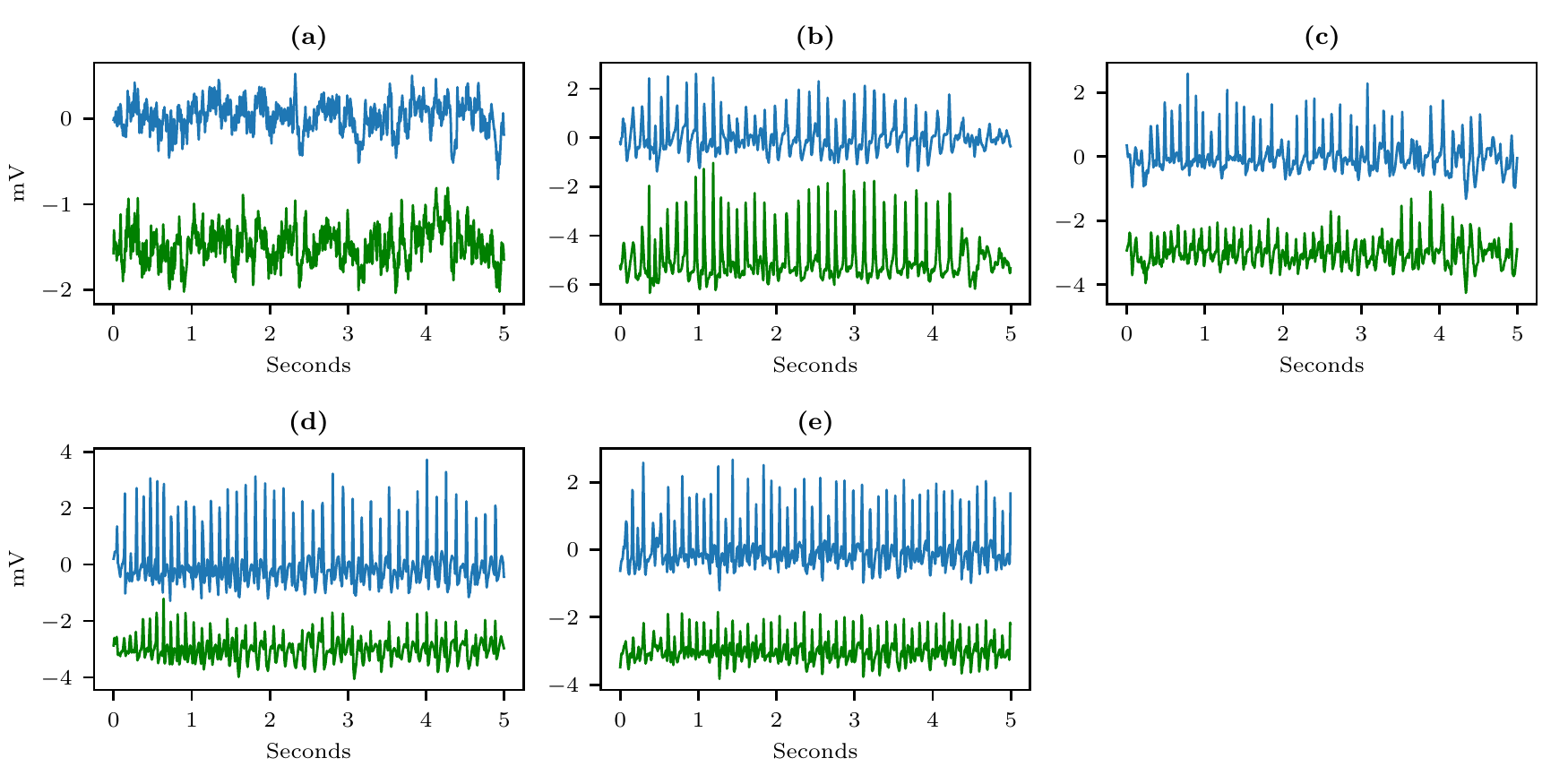}
    \caption{The five considered examples of rat EEGs, with (a) a normal non-epileptic signal and (b)-(e) epileptic spike trains. Left hemisphere signals (green) are plotted with a vertical offset (\SI{-1.5}{\mV} in (a), \SI{-5}{\mV} in (b), and \SI{-3}{\mV} in (c)-(e)) for better visibility, whereas right hemisphere signals (blue) are plotted without an offset. Note the different vertical scales.}
    \label{fig:rat_plot}
\end{figure*}

As a second rather common example for the application of bivariate event based statistics, we analyze five pairs of rat EEG signals obtained from electrodes at the left and right frontal cortex of male adult rats, including one non-epileptic case and four epileptic cases. This dataset was also included in \textcite{QuianQuiroga2002} and is publicly available \footnote{The rat EEG data may be downloaded from \url{https://www2.le.ac.uk/centres/csn/software} as dataset 4.} with more detailed information on experiment settings, recordings and results to be found in \cite{Luijtelaar1997}. Each signal was recorded for \SI{5}{\second} and digitized at \SI{200}{\hertz}, yielding a time series of 1,000 values. Fig.~\ref{fig:rat_plot} shows plots of the five EEG pairs. Example (a) displays the normal non-epileptic signal, while examples (b) to (e) all exhibit clear epileptic activity as apparent from regular spike discharges.

Under the plausible assumption that neither the spike shape nor the background noise carry valuable information, we may again extract events from the depicted time series, for which we compare two options. Firstly, similar to the climate data case, we simply impose a threshold at the 97th percentile to obtain a well-defined discrimination of extreme events. Secondly, we follow the approach of \textcite{QuianQuiroga2002} by declaring events at $t$ as local maxima fulfilling the following conditions: (1) $x_t > x_{t+k}$ for $k=-K,\dotsc,-1,1,\dotsc,K$ (also including $k=0$ as stated in the original reference appears incorrect since $x_t>x_t$ would never be satisfied) and (2) $x_t > x_{t \pm K} + h$, with free parameters $K$ and $h$. In agreement with the original definition, we set $K=3$ and $h=0.1$. Subsequently, we calculate $Q_{ij}^{ES}$ and $Q_{ij}^{ECA,mean}$ for both types of event definitions. 

Our results do not exactly replicate those in \textcite{QuianQuiroga2002} as they replaced the dynamic local ES coincidence interval $\tau_{lm}^{ij}$ with a fixed global value of 2 time steps, i.e., \SI{10}{\ms}. In our opinion, this practice of educated guessing should be handled with care as it might void the normalization by enabling unintentional double counting of event pairs. However, to allow for some degree of comparison, we set $\Delta T = \SI{10}{\ms}$ and $\tau = \SI{0}{ms}$ for the ECA parameters, which can be done without jeopardizing normalization.

Table~\ref{tab:rat_es_eca} displays the results of ES and ECA for both event definitions and all five EEG pairs. Since ES and ECA merely share a normalization to the same interval but differ substantially between these boundaries, absolute values are not strictly comparable as also mentioned previously. In order to allow for a fair comparison, we thus also provide the percentage values in one column as compared with the maximum value of that column. Note that this relative approach is also consistent with our strategy used when studying the artificial simulation and the real world climate data. In the first case relative patterns rather than absolute values were analyzed, while in the second case the network construction process extracted only the strongest links on a relative basis (see Fig.~\ref{fig:climnet}).

\begin{table}
    \caption{\label{tab:rat_es_eca}ES and ECA results for five selected rat EEGs.}
    \begin{ruledtabular}
    \begin{tabular}{lllll}
    & \multicolumn{2}{c}{Threshold exceedance} & \multicolumn{2}{c}{Method from \cite{QuianQuiroga2002}} \\ 
    Case & $Q_{ij}^{ES}$ & $Q_{ij}^{ECA,mean}$ & $Q_{ij}^{ES}$ & $Q_{ij}^{ECA,mean}$ \\
    (a) & 0.22 (31\%)  & 0.32 (43\%)  & 0.70 (78\%)  & 0.55 (79\%)  \\
    (b) & 0.50 (70\%)  & 0.63 (85\%)  & 0.77 (85\%)  & 0.53 (76\%)  \\
    (c) & 0.21 (30\%)  & 0.23 (31\%)  & 0.77 (85\%)  & 0.51 (74\%)  \\
    (d) & 0.43 (60\%)  & 0.45 (60\%)  & 0.85 (94\%)  & 0.57 (81\%)  \\
    (e) & 0.71 (100\%) & 0.75 (100\%) & 0.91 (100\%) & 0.70 (100\%) \\
    \end{tabular}
    \end{ruledtabular}
\end{table}

In relative terms, the differences between ES and ECA turn out to be small for both event definitions, with a maximum difference of 15\% for example (b). All other examples exhibit even smaller differences. Irrespective of either the event definition or the similarity measure, within the epileptic signals (e) is always ranked first (most strongly synchronized) and (c) last (least synchronized). Within each event definition method and again except for example (a), the ranking order remains consistent across both ES and ECA. This confirms that the results of both approaches resemble each other fairly well for time series that are characterized by relatively constant event rates. However, this observation even holds true for the non-epileptic case (a), where results are also comparable for both event definitions. Yet, we do observe pronounced differences between the two event definitions, which are discussed in more detail below. Of course, the ECA results depend markedly on the parameter values and similar values can be obtained by setting the delay $\tau$ sufficiently close to the mean inter-event distance. However, admittedly, the potential to analyze the temporal evolution of event synchrony might not be regarded as an equally important feature for EEG as for climate data.

Thus, if the two different event definitions are considered separately, ES and ECA yield very similar results. This is a direct and plausible consequence of a sufficiently narrow inter-event distance distribution for epileptic EEG spike trains in examples (b)-(e), which stands in marked contrast to the precipitation data used previously. Since ES was originally designed with EEG applications in mind, our findings are conceptually justified and not surprising. While the observed consistency of ES and ECA was also facilitated by very distinct events due to the recorded epileptic activity in (b)-(e), notably this also holds true for the non-epileptic example (a), most likely because the probability distribution of waiting times between subsequent events hardly shows any values close to zero (not shown), which in turn would correspond to a regime where we might expect deviations between the two methods. This finding further underlines the versatility of ECA.

The importance of unambiguous event extraction is moreover revealed in examples (a) and (c). In example (a), the threshold method only yields results of 0.22 (31\%) and 0.32 (43\%), while the event definition from \cite{QuianQuiroga2002} leads to values of 0.70 (78\%) and 0.55 (79\%). Similarly, in example (c) we obtain 0.21 (30\%) and 0.23 (31\%) using threshold events versus 0.77 (85\%) and 0.51 (74\%) using \cite{QuianQuiroga2002}. In both examples, these differences in both absolute, but more importantly also relative values, are very likely caused by less pronounced peaks over a dynamic background in (a) and (c) as compared with the other examples, which probably led to error-prone event definitions for a simple threshold method. This already hints to the overarching issue of statistically disentangling events from underlying phenomena, which will be further discussed below in conjunction with the subsequent choice of a proper similarity measure.

\section{Discussion \label{sec:discussion}}

Within the scope of (extreme) event analysis, the problem of serial dependencies in time can principally be tackled in two ways: either by choosing a robust analysis method or by defining events in different ways. For the first approach, we have provided here tentative explanations why ECA may outperform ES. However, it appears also viable to address the issue already in the previous step of event definition. This necessarily requires more involved preprocessing techniques than a mere threshold exceedance strategy.

For EEG signals, the temporal resolution relative to the number of spikes is usually much higher than for climatological data. Thus, several time values that clearly belong to the same peak might fulfill the threshold exceedance criterion, even for very high percentiles. Since the focus in EEG spike train analysis is on the extraction of singular events that are \textit{local} (or relative) maxima, which may well have different amplitudes among themselves, the event definition method by \textcite{QuianQuiroga2002} is a sensible approach for this delicate task. For climatological extremes, we are however not interested in local, but indeed rather in \textit{global} (or absolute) maxima with respect to some threshold (i.e., taking a peaks-over-threshold approach as common in extreme value theory) because these are the type of events with potentially devastating impacts. Thus, applying said EEG event definition method also to climate data appears unreasonable. Yet another possibility would be the utilization of sophisticated clustering algorithms. However, we reject this as unnecessarily complex for the sake of the present work, since ECA fulfills the same purpose in a much more straightforward and easily understandable manner.

In a broader context, the task of disentangling event synchrony from serial dependency therefore transitions into the more profound endeavour of disentangling statistical events from underlying phenomena, which we are eventually interested in. In our opinion, this must be either informed by a priori knowledge of the system or guided by specific research questions targeted to the given dataset under study. A shared feature of the two aforementioned options is that they both require some external parameters, which determine the expected time scales of serial dependencies and which cannot be set independent of data and context. In essence, the values $K$ and $h$ for the event definition in \cite{QuianQuiroga2002} or $\Delta T$ and $\tau$ for ECA constitute different parameterizations of just this issue. Thus, in our opinion, the introduction of a sophisticated clustering algorithm \cite{Boers2014}, which merges several previously defined global extremes into one, may only shift the choice of these time scale parameters elsewhere, without relieving us of the actual task of determining them.

In this line of argument, there seems to be no universally optimal method. While the inclusion of multiple time scales into one single association measure for event sequences might constitute a noteworthy advantage of ES over ECA, this can only be truly exploited if the data has been preprocessed diligently. Succumbing to the tempting pitfall of using ES without preprocessing makes it vulnerable to the identified weaknesses originating from serial dependencies among events. ECA, on the other hand, offers a more refined analysis of time scales through $\Delta T$ (and $\tau$) with the considerable advantage of dealing properly with both serially dependent and independent event time series, albeit without the possibility to dynamically incorporate changing event rates into a single measure. Using a sliding windows approach could however alleviate this alleged restriction.

For future research on functional climate networks, we therefore reinforce our suggestion to use ECA instead of ES. Even if serial dependencies were dealt with before so that ES worked as intended, the algorithm would still lack a clear declaration on the involved time scales as well as the possibility to scrutinize the temporal evolution, which we perceive as a valuable advantage in its own right. Additionally, using ECA elegantly circumvents the need for declustering along with its ensuing parameterization difficulties. Only if the inclusion of varying time scales into one single measure is essential, if events are clearly delineated and if refined temporal analyses are undesired, ES appears to be the method of choice. This is probably the case for EEG data, even though ECA can also be adjusted to fit such applications as well.

At a final note, we emphasize that many other measures have been used in the past for quantifying different aspects of statistical interdependence between two time series, particularly in the context of functional network analysis. For example, in the context of spatiotemporal fields of climate data, Pearson correlation and mutual information have been often used for this purpose. To understand the differences with respect to ES and ECA used in the present work, one should note that those (as well as many other) measures take \emph{all} existing data points (i.e., values from both the bulk and the tails of the probability distributions of the variable of interest) into account and attempt to quantify the strength of a linear (Pearson correlation) or arbitrary functional relationship (mutual information) between two series. Hence, those measures are necessarily dominated by statistical associations among the bulk values due to their by far larger number than that of the values in the tails. In turn, there are important applications where statistical associations among those bulk values are not of primary interest, since the relationship between extraordinary values (or extremes) can be believed to differ from that under ``normal conditions'' \cite{Siegmund2016a} (for example, in the precipitation example discussed in Sec.~\ref{subsec:climdata}). On purpose, only the latter aspect has been addressed in the present work. Due to their conceptual difference (continuum-based versus event-based association measures), inferred statistical associations among the given data sets may crucially change when either considering all data or focusing only on the extremes (or even more, in our present work, only the timing of extremes and not their magnitudes). Inter-comparisons between measures of both types have been published elsewhere (e.g., in \cite{QuianQuiroga2002a} for neuroscience applications or \cite{Ferreira2019} for climate data), and we did not attempt to repeat such studies here to maintain the topical focus of the present work.

\section{Conclusions \label{sec:conclusion}}

In this paper, we have explored the key differences of two statistical association measures for event time series, event synchronization (ES) and event coincidence analysis (ECA). Both measures have been used extensively in different disciplines, yet had hardly been systematically compared or applied to the same data before \cite{Hassanibesheli2019}.

First of all, building on identified ambiguities in the theoretical definitions of both measures, we introduced and implemented corrected versions of the original proposals. We then created artificial data from a coupled autoregressive process, by which we were able to provide evidence that ES does not allow to unambiguously disentangle event synchrony from serial dependencies, whereas ECA is significantly less susceptible to such issues. Reproducing the results of previous studies, we demonstrated ensuing implications for real-world data that reinforce our conceptual concerns. We specifically provided indications that results from several past climate network studies, which rely on ES as a similarity metric, need to be interpreted with caution as they might exhibit severe biases originating from unaddressed serial dependencies between events. On the other hand, for epileptic EEG data, we showed that both ES and ECA yield qualitatively similar results as the characteristic spike trains entail relatively constant event rates without notable temporal clustering.

Furthermore, we discussed the nexus of event definitions and appropriate similarity measures conceptually. We argued that disentangling event synchrony from serial dependency is on a lower level tantamount to disentangling statistical events from underlying phenomena. While methods that extract local extremes prove to be sensible for data with clear spikes of varying amplitude such as EEG signals, they are not applicable for cases where the focus lies on global maxima such as extreme climate events. Because ES only works properly for preprocessed data with a priori knowledge, we propose ECA as an alternative measure that can handle both serially dependent and independent data. Furthermore, ECA allows to explicitly analyze temporal evolution and elegantly bypasses the need for unnecessarily complex clustering algorithms that would be required if we wanted to analyze extremes with ES.

While both ES and ECA have strength and weaknesses, the nonparametric nature of ES makes it all too easy to succumb to the temptation of omitting the definition of a time scale, within which multiple events belong to the same phenomenon. However, it appears impossible to leap over this step for synchronization to be attributed between meaningfully defined events. Whether it is best to stipulate this time scale via parameters in the event definition or rather in the subsequent analysis, remains open for debate at this point. Even though the quest for a universally optimal method in our view thus constitutes an ill-posed problem, we advocate in favor of ECA as providing a straightforward event based statistical association measure to analyze event time series that may or may not include serial dependencies, without caveats due to temporal event clustering.

Ultimately, the question which event definition and which similarity measure is most appropriate remains a matter of choice. But that choice should be made well-informed, imperatively embedded into the research context and data characteristics.

\begin{acknowledgments}
This work has been financially supported by the German Federal Ministry for Education and Research via the Young Investigators Group CoSy-CC2: Complex Systems Approaches to Understanding Causes and Consequences of Past, Present and Future Climate Change (grant no. 01LN1306A) and the Belmont Forum/JPI Climate project GOTHAM (Globally Observed Teleconnections and their Representation in Hierarchies of Atmospheric Models, grant no. 01LP16MA). AO has been partially funded by the Foundation of German Business (sdw) and received travel support from the Graduate School of Integrated Climate System Sciences (SICSS) of the Cluster of Excellence CLICCS at the University of Hamburg. AO would like to express his gratitude to Joachim Krug for supporting and co-supervising parts of the presented work at the University of Cologne. Functional network analysis has been performed using the Python package \texttt{pyunicorn} \cite{Donges2015} freely available at GitHub.
\end{acknowledgments}

\bibliography{bib_03.bib}

\begin{thebibliography}{51}%
\makeatletter
\providecommand \@ifxundefined [1]{%
 \@ifx{#1\undefined}
}%
\providecommand \@ifnum [1]{%
 \ifnum #1\expandafter \@firstoftwo
 \else \expandafter \@secondoftwo
 \fi
}%
\providecommand \@ifx [1]{%
 \ifx #1\expandafter \@firstoftwo
 \else \expandafter \@secondoftwo
 \fi
}%
\providecommand \natexlab [1]{#1}%
\providecommand \enquote  [1]{``#1''}%
\providecommand \bibnamefont  [1]{#1}%
\providecommand \bibfnamefont [1]{#1}%
\providecommand \citenamefont [1]{#1}%
\providecommand \href@noop [0]{\@secondoftwo}%
\providecommand \href [0]{\begingroup \@sanitize@url \@href}%
\providecommand \@href[1]{\@@startlink{#1}\@@href}%
\providecommand \@@href[1]{\endgroup#1\@@endlink}%
\providecommand \@sanitize@url [0]{\catcode `\\12\catcode `\$12\catcode
  `\&12\catcode `\#12\catcode `\^12\catcode `\_12\catcode `\%12\relax}%
\providecommand \@@startlink[1]{}%
\providecommand \@@endlink[0]{}%
\providecommand \url  [0]{\begingroup\@sanitize@url \@url }%
\providecommand \@url [1]{\endgroup\@href {#1}{\urlprefix }}%
\providecommand \urlprefix  [0]{URL }%
\providecommand \Eprint [0]{\href }%
\providecommand \doibase [0]{http://dx.doi.org/}%
\providecommand \selectlanguage [0]{\@gobble}%
\providecommand \bibinfo  [0]{\@secondoftwo}%
\providecommand \bibfield  [0]{\@secondoftwo}%
\providecommand \translation [1]{[#1]}%
\providecommand \BibitemOpen [0]{}%
\providecommand \bibitemStop [0]{}%
\providecommand \bibitemNoStop [0]{.\EOS\space}%
\providecommand \EOS [0]{\spacefactor3000\relax}%
\providecommand \BibitemShut  [1]{\csname bibitem#1\endcsname}%
\let\auto@bib@innerbib\@empty
\bibitem [{\citenamefont {Albeverio}\ \emph {et~al.}(2006)\citenamefont
  {Albeverio}, \citenamefont {Jentsch},\ and\ \citenamefont
  {Kantz}}]{Albeverio2006}%
  \BibitemOpen
  \bibfield  {author} {\bibinfo {author} {\bibfnamefont {S.}~\bibnamefont
  {Albeverio}}, \bibinfo {author} {\bibfnamefont {V.}~\bibnamefont {Jentsch}},
  \ and\ \bibinfo {author} {\bibfnamefont {H.}~\bibnamefont {Kantz}},\
  }\href@noop {} {\emph {\bibinfo {title} {Extreme events in nature and
  society}}}\ (\bibinfo  {publisher} {Springer},\ \bibinfo {address} {Berlin},\
  \bibinfo {year} {2006})\BibitemShut {NoStop}%
\bibitem [{\citenamefont {Quian~Quiroga}\ \emph
  {et~al.}(2002{\natexlab{a}})\citenamefont {Quian~Quiroga}, \citenamefont
  {Kreuz},\ and\ \citenamefont {Grassberger}}]{QuianQuiroga2002}%
  \BibitemOpen
  \bibfield  {author} {\bibinfo {author} {\bibfnamefont {R.}~\bibnamefont
  {Quian~Quiroga}}, \bibinfo {author} {\bibfnamefont {T.}~\bibnamefont
  {Kreuz}}, \ and\ \bibinfo {author} {\bibfnamefont {P.}~\bibnamefont
  {Grassberger}},\ }\href {\doibase 10.1103/PhysRevE.66.041904} {\bibfield
  {journal} {\bibinfo  {journal} {Physical Review E}\ }\textbf {\bibinfo
  {volume} {66}},\ \bibinfo {pages} {041904} (\bibinfo {year}
  {2002}{\natexlab{a}})}\BibitemShut {NoStop}%
\bibitem [{\citenamefont {Donges}\ \emph {et~al.}(2011)\citenamefont {Donges},
  \citenamefont {Donner}, \citenamefont {Trauth}, \citenamefont {Marwan},
  \citenamefont {Schellnhuber},\ and\ \citenamefont {Kurths}}]{Donges2011}%
  \BibitemOpen
  \bibfield  {author} {\bibinfo {author} {\bibfnamefont {J.~F.}\ \bibnamefont
  {Donges}}, \bibinfo {author} {\bibfnamefont {R.~V.}\ \bibnamefont {Donner}},
  \bibinfo {author} {\bibfnamefont {M.~H.}\ \bibnamefont {Trauth}}, \bibinfo
  {author} {\bibfnamefont {N.}~\bibnamefont {Marwan}}, \bibinfo {author}
  {\bibfnamefont {H.-J.}\ \bibnamefont {Schellnhuber}}, \ and\ \bibinfo
  {author} {\bibfnamefont {J.}~\bibnamefont {Kurths}},\ }\href@noop {}
  {\bibfield  {journal} {\bibinfo  {journal} {Proceedings of the National
  Academy of Sciences}\ }\textbf {\bibinfo {volume} {108}},\ \bibinfo {pages}
  {20422} (\bibinfo {year} {2011})}\BibitemShut {NoStop}%
\bibitem [{\citenamefont {Donges}\ \emph {et~al.}(2016)\citenamefont {Donges},
  \citenamefont {Schleussner}, \citenamefont {Siegmund},\ and\ \citenamefont
  {Donner}}]{Donges2016}%
  \BibitemOpen
  \bibfield  {author} {\bibinfo {author} {\bibfnamefont {J.}~\bibnamefont
  {Donges}}, \bibinfo {author} {\bibfnamefont {C.-F.}\ \bibnamefont
  {Schleussner}}, \bibinfo {author} {\bibfnamefont {J.}~\bibnamefont
  {Siegmund}}, \ and\ \bibinfo {author} {\bibfnamefont {R.}~\bibnamefont
  {Donner}},\ }\href {\doibase 10.1140/epjst/e2015-50233-y} {\bibfield
  {journal} {\bibinfo  {journal} {European Physical Journal Special Topics}\
  }\textbf {\bibinfo {volume} {225}},\ \bibinfo {pages} {471} (\bibinfo {year}
  {2016})}\BibitemShut {NoStop}%
\bibitem [{\citenamefont {Kreuz}\ \emph {et~al.}(2004)\citenamefont {Kreuz},
  \citenamefont {Andrzejak}, \citenamefont {Mormann}, \citenamefont {Kraskov},
  \citenamefont {St\"ogbauer}, \citenamefont {Elger}, \citenamefont
  {Lehnertz},\ and\ \citenamefont {Grassberger}}]{Kreuz2004}%
  \BibitemOpen
  \bibfield  {author} {\bibinfo {author} {\bibfnamefont {T.}~\bibnamefont
  {Kreuz}}, \bibinfo {author} {\bibfnamefont {R.~G.}\ \bibnamefont
  {Andrzejak}}, \bibinfo {author} {\bibfnamefont {F.}~\bibnamefont {Mormann}},
  \bibinfo {author} {\bibfnamefont {A.}~\bibnamefont {Kraskov}}, \bibinfo
  {author} {\bibfnamefont {H.}~\bibnamefont {St\"ogbauer}}, \bibinfo {author}
  {\bibfnamefont {C.~E.}\ \bibnamefont {Elger}}, \bibinfo {author}
  {\bibfnamefont {K.}~\bibnamefont {Lehnertz}}, \ and\ \bibinfo {author}
  {\bibfnamefont {P.}~\bibnamefont {Grassberger}},\ }\href {\doibase
  10.1103/PhysRevE.69.061915} {\bibfield  {journal} {\bibinfo  {journal}
  {Physical Review E}\ }\textbf {\bibinfo {volume} {69}},\ \bibinfo {pages}
  {061915} (\bibinfo {year} {2004})}\BibitemShut {NoStop}%
\bibitem [{\citenamefont {Pereda}\ \emph {et~al.}(2005)\citenamefont {Pereda},
  \citenamefont {Quiroga},\ and\ \citenamefont {Bhattacharya}}]{Pereda2005}%
  \BibitemOpen
  \bibfield  {author} {\bibinfo {author} {\bibfnamefont {E.}~\bibnamefont
  {Pereda}}, \bibinfo {author} {\bibfnamefont {R.~Q.}\ \bibnamefont {Quiroga}},
  \ and\ \bibinfo {author} {\bibfnamefont {J.}~\bibnamefont {Bhattacharya}},\
  }\href {\doibase https://doi.org/10.1016/j.pneurobio.2005.10.003} {\bibfield
  {journal} {\bibinfo  {journal} {Progress in Neurobiology}\ }\textbf {\bibinfo
  {volume} {77}},\ \bibinfo {pages} {1 } (\bibinfo {year} {2005})}\BibitemShut
  {NoStop}%
\bibitem [{\citenamefont {Kreuz}\ \emph
  {et~al.}(2007{\natexlab{a}})\citenamefont {Kreuz}, \citenamefont {Haas},
  \citenamefont {Morelli}, \citenamefont {Abarbanel},\ and\ \citenamefont
  {Politi}}]{Kreuz2007}%
  \BibitemOpen
  \bibfield  {author} {\bibinfo {author} {\bibfnamefont {T.}~\bibnamefont
  {Kreuz}}, \bibinfo {author} {\bibfnamefont {J.~S.}\ \bibnamefont {Haas}},
  \bibinfo {author} {\bibfnamefont {A.}~\bibnamefont {Morelli}}, \bibinfo
  {author} {\bibfnamefont {H.~D.}\ \bibnamefont {Abarbanel}}, \ and\ \bibinfo
  {author} {\bibfnamefont {A.}~\bibnamefont {Politi}},\ }\href {\doibase
  https://doi.org/10.1016/j.jneumeth.2007.05.031} {\bibfield  {journal}
  {\bibinfo  {journal} {Journal of Neuroscience Methods}\ }\textbf {\bibinfo
  {volume} {165}},\ \bibinfo {pages} {151 } (\bibinfo {year}
  {2007}{\natexlab{a}})}\BibitemShut {NoStop}%
\bibitem [{\citenamefont {Kreuz}\ \emph
  {et~al.}(2007{\natexlab{b}})\citenamefont {Kreuz}, \citenamefont {Mormann},
  \citenamefont {Andrzejak}, \citenamefont {Kraskov}, \citenamefont
  {Lehnertz},\ and\ \citenamefont {Grassberger}}]{Kreuz2007a}%
  \BibitemOpen
  \bibfield  {author} {\bibinfo {author} {\bibfnamefont {T.}~\bibnamefont
  {Kreuz}}, \bibinfo {author} {\bibfnamefont {F.}~\bibnamefont {Mormann}},
  \bibinfo {author} {\bibfnamefont {R.~G.}\ \bibnamefont {Andrzejak}}, \bibinfo
  {author} {\bibfnamefont {A.}~\bibnamefont {Kraskov}}, \bibinfo {author}
  {\bibfnamefont {K.}~\bibnamefont {Lehnertz}}, \ and\ \bibinfo {author}
  {\bibfnamefont {P.}~\bibnamefont {Grassberger}},\ }\href {\doibase
  https://doi.org/10.1016/j.physd.2006.09.039} {\bibfield  {journal} {\bibinfo
  {journal} {Physica D: Nonlinear Phenomena}\ }\textbf {\bibinfo {volume}
  {225}},\ \bibinfo {pages} {29 } (\bibinfo {year}
  {2007}{\natexlab{b}})}\BibitemShut {NoStop}%
\bibitem [{\citenamefont {Stam}(2005)}]{Stam2005}%
  \BibitemOpen
  \bibfield  {author} {\bibinfo {author} {\bibfnamefont {C.}~\bibnamefont
  {Stam}},\ }\href {\doibase https://doi.org/10.1016/j.clinph.2005.06.011}
  {\bibfield  {journal} {\bibinfo  {journal} {Clinical Neurophysiology}\
  }\textbf {\bibinfo {volume} {116}},\ \bibinfo {pages} {2266 } (\bibinfo
  {year} {2005})}\BibitemShut {NoStop}%
\bibitem [{\citenamefont {Dauwels}\ \emph {et~al.}(2008)\citenamefont
  {Dauwels}, \citenamefont {Vialatte}, \citenamefont {Weber},\ and\
  \citenamefont {Cichocki}}]{Dauwels2008}%
  \BibitemOpen
  \bibfield  {author} {\bibinfo {author} {\bibfnamefont {J.}~\bibnamefont
  {Dauwels}}, \bibinfo {author} {\bibfnamefont {F.}~\bibnamefont {Vialatte}},
  \bibinfo {author} {\bibfnamefont {T.}~\bibnamefont {Weber}}, \ and\ \bibinfo
  {author} {\bibfnamefont {A.}~\bibnamefont {Cichocki}},\ }in\ \href@noop {}
  {\emph {\bibinfo {booktitle} {International Conference on Neural Information
  Processing}}}\ (\bibinfo {organization} {Springer},\ \bibinfo {year} {2008})\
  pp.\ \bibinfo {pages} {177--185}\BibitemShut {NoStop}%
\bibitem [{\citenamefont {Angotzi}\ \emph {et~al.}(2014)\citenamefont
  {Angotzi}, \citenamefont {Boi}, \citenamefont {Zordan}, \citenamefont
  {Bonfanti},\ and\ \citenamefont {Vato}}]{Angotzi2014}%
  \BibitemOpen
  \bibfield  {author} {\bibinfo {author} {\bibfnamefont {G.~N.}\ \bibnamefont
  {Angotzi}}, \bibinfo {author} {\bibfnamefont {F.}~\bibnamefont {Boi}},
  \bibinfo {author} {\bibfnamefont {S.}~\bibnamefont {Zordan}}, \bibinfo
  {author} {\bibfnamefont {A.}~\bibnamefont {Bonfanti}}, \ and\ \bibinfo
  {author} {\bibfnamefont {A.}~\bibnamefont {Vato}},\ }\href@noop {} {\bibfield
   {journal} {\bibinfo  {journal} {Scientific Reports}\ }\textbf {\bibinfo
  {volume} {4}},\ \bibinfo {pages} {5963} (\bibinfo {year} {2014})}\BibitemShut
  {NoStop}%
\bibitem [{\citenamefont {Singh}\ \emph {et~al.}(2014)\citenamefont {Singh},
  \citenamefont {Gibbons}, \citenamefont {Saravanaperumal}, \citenamefont {Du},
  \citenamefont {Hennig}, \citenamefont {Eisenman}, \citenamefont {Mazzone},
  \citenamefont {Hayashi}, \citenamefont {Cao}, \citenamefont {Stoltz} \emph
  {et~al.}}]{Singh2014}%
  \BibitemOpen
  \bibfield  {author} {\bibinfo {author} {\bibfnamefont {R.~D.}\ \bibnamefont
  {Singh}}, \bibinfo {author} {\bibfnamefont {S.~J.}\ \bibnamefont {Gibbons}},
  \bibinfo {author} {\bibfnamefont {S.~A.}\ \bibnamefont {Saravanaperumal}},
  \bibinfo {author} {\bibfnamefont {P.}~\bibnamefont {Du}}, \bibinfo {author}
  {\bibfnamefont {G.~W.}\ \bibnamefont {Hennig}}, \bibinfo {author}
  {\bibfnamefont {S.~T.}\ \bibnamefont {Eisenman}}, \bibinfo {author}
  {\bibfnamefont {A.}~\bibnamefont {Mazzone}}, \bibinfo {author} {\bibfnamefont
  {Y.}~\bibnamefont {Hayashi}}, \bibinfo {author} {\bibfnamefont
  {C.}~\bibnamefont {Cao}}, \bibinfo {author} {\bibfnamefont {G.~J.}\
  \bibnamefont {Stoltz}},  \emph {et~al.},\ }\href@noop {} {\bibfield
  {journal} {\bibinfo  {journal} {Journal of Physiology}\ }\textbf {\bibinfo
  {volume} {592}},\ \bibinfo {pages} {4051} (\bibinfo {year}
  {2014})}\BibitemShut {NoStop}%
\bibitem [{\citenamefont {Varni}\ \emph {et~al.}(2010)\citenamefont {Varni},
  \citenamefont {Volpe},\ and\ \citenamefont {Camurri}}]{Varni2010}%
  \BibitemOpen
  \bibfield  {author} {\bibinfo {author} {\bibfnamefont {G.}~\bibnamefont
  {Varni}}, \bibinfo {author} {\bibfnamefont {G.}~\bibnamefont {Volpe}}, \ and\
  \bibinfo {author} {\bibfnamefont {A.}~\bibnamefont {Camurri}},\ }\href@noop
  {} {\bibfield  {journal} {\bibinfo  {journal} {IEEE Transactions on
  Multimedia}\ }\textbf {\bibinfo {volume} {12}},\ \bibinfo {pages} {576}
  (\bibinfo {year} {2010})}\BibitemShut {NoStop}%
\bibitem [{\citenamefont {Butail}\ \emph {et~al.}(2016)\citenamefont {Butail},
  \citenamefont {Mwaffo},\ and\ \citenamefont {Porfiri}}]{Butail2016}%
  \BibitemOpen
  \bibfield  {author} {\bibinfo {author} {\bibfnamefont {S.}~\bibnamefont
  {Butail}}, \bibinfo {author} {\bibfnamefont {V.}~\bibnamefont {Mwaffo}}, \
  and\ \bibinfo {author} {\bibfnamefont {M.}~\bibnamefont {Porfiri}},\
  }\href@noop {} {\bibfield  {journal} {\bibinfo  {journal} {Physical Review
  E}\ }\textbf {\bibinfo {volume} {93}},\ \bibinfo {pages} {042411} (\bibinfo
  {year} {2016})}\BibitemShut {NoStop}%
\bibitem [{\citenamefont {Zhou}\ and\ \citenamefont
  {Sornette}(2003)}]{Zhou2003}%
  \BibitemOpen
  \bibfield  {author} {\bibinfo {author} {\bibfnamefont {W.-X.}\ \bibnamefont
  {Zhou}}\ and\ \bibinfo {author} {\bibfnamefont {D.}~\bibnamefont
  {Sornette}},\ }\href@noop {} {\bibfield  {journal} {\bibinfo  {journal}
  {Physica A: Statistical Mechanics and its Applications}\ }\textbf {\bibinfo
  {volume} {330}},\ \bibinfo {pages} {543} (\bibinfo {year}
  {2003})}\BibitemShut {NoStop}%
\bibitem [{\citenamefont {Malik}\ \emph {et~al.}(2010)\citenamefont {Malik},
  \citenamefont {Marwan},\ and\ \citenamefont {Kurths}}]{Malik2010}%
  \BibitemOpen
  \bibfield  {author} {\bibinfo {author} {\bibfnamefont {N.}~\bibnamefont
  {Malik}}, \bibinfo {author} {\bibfnamefont {N.}~\bibnamefont {Marwan}}, \
  and\ \bibinfo {author} {\bibfnamefont {J.}~\bibnamefont {Kurths}},\ }\href
  {\doibase 10.5194/npg-17-371-2010} {\bibfield  {journal} {\bibinfo  {journal}
  {Nonlinear Processes in Geophysics}\ }\textbf {\bibinfo {volume} {17}},\
  \bibinfo {pages} {371} (\bibinfo {year} {2010})}\BibitemShut {NoStop}%
\bibitem [{\citenamefont {Malik}\ \emph {et~al.}(2012)\citenamefont {Malik},
  \citenamefont {Bookhagen}, \citenamefont {Marwan},\ and\ \citenamefont
  {Kurths}}]{Malik2012}%
  \BibitemOpen
  \bibfield  {author} {\bibinfo {author} {\bibfnamefont {N.}~\bibnamefont
  {Malik}}, \bibinfo {author} {\bibfnamefont {B.}~\bibnamefont {Bookhagen}},
  \bibinfo {author} {\bibfnamefont {N.}~\bibnamefont {Marwan}}, \ and\ \bibinfo
  {author} {\bibfnamefont {J.}~\bibnamefont {Kurths}},\ }\href {\doibase
  10.1007/s00382-011-1156-4} {\bibfield  {journal} {\bibinfo  {journal}
  {Climate Dynamics}\ }\textbf {\bibinfo {volume} {39}},\ \bibinfo {pages}
  {971} (\bibinfo {year} {2012})}\BibitemShut {NoStop}%
\bibitem [{\citenamefont {Boers}\ \emph {et~al.}(2013)\citenamefont {Boers},
  \citenamefont {Bookhagen}, \citenamefont {Marwan}, \citenamefont {Kurths},\
  and\ \citenamefont {Marengo}}]{Boers2013}%
  \BibitemOpen
  \bibfield  {author} {\bibinfo {author} {\bibfnamefont {N.}~\bibnamefont
  {Boers}}, \bibinfo {author} {\bibfnamefont {B.}~\bibnamefont {Bookhagen}},
  \bibinfo {author} {\bibfnamefont {N.}~\bibnamefont {Marwan}}, \bibinfo
  {author} {\bibfnamefont {J.}~\bibnamefont {Kurths}}, \ and\ \bibinfo {author}
  {\bibfnamefont {J.}~\bibnamefont {Marengo}},\ }\href {\doibase
  10.1002/grl.50681} {\bibfield  {journal} {\bibinfo  {journal} {Geophysical
  Research Letters}\ }\textbf {\bibinfo {volume} {40}},\ \bibinfo {pages}
  {4386} (\bibinfo {year} {2013})}\BibitemShut {NoStop}%
\bibitem [{\citenamefont {Boers}\ \emph
  {et~al.}(2014{\natexlab{a}})\citenamefont {Boers}, \citenamefont {Bookhagen},
  \citenamefont {Barbosa}, \citenamefont {Marwan}, \citenamefont {Kurths},\
  and\ \citenamefont {Marengo}}]{Boers2014}%
  \BibitemOpen
  \bibfield  {author} {\bibinfo {author} {\bibfnamefont {N.}~\bibnamefont
  {Boers}}, \bibinfo {author} {\bibfnamefont {B.}~\bibnamefont {Bookhagen}},
  \bibinfo {author} {\bibfnamefont {H.~M.~J.}\ \bibnamefont {Barbosa}},
  \bibinfo {author} {\bibfnamefont {N.}~\bibnamefont {Marwan}}, \bibinfo
  {author} {\bibfnamefont {J.}~\bibnamefont {Kurths}}, \ and\ \bibinfo {author}
  {\bibfnamefont {J.~A.}\ \bibnamefont {Marengo}},\ }\href {\doibase
  10.1038/ncomms6199} {\bibfield  {journal} {\bibinfo  {journal} {Nature
  Communications}\ }\textbf {\bibinfo {volume} {5}},\ \bibinfo {pages} {5199}
  (\bibinfo {year} {2014}{\natexlab{a}})}\BibitemShut {NoStop}%
\bibitem [{\citenamefont {Boers}\ \emph
  {et~al.}(2014{\natexlab{b}})\citenamefont {Boers}, \citenamefont {Rheinwalt},
  \citenamefont {Bookhagen}, \citenamefont {Barbosa}, \citenamefont {Marwan},
  \citenamefont {Marengo},\ and\ \citenamefont {Kurths}}]{Boers2014a}%
  \BibitemOpen
  \bibfield  {author} {\bibinfo {author} {\bibfnamefont {N.}~\bibnamefont
  {Boers}}, \bibinfo {author} {\bibfnamefont {A.}~\bibnamefont {Rheinwalt}},
  \bibinfo {author} {\bibfnamefont {B.}~\bibnamefont {Bookhagen}}, \bibinfo
  {author} {\bibfnamefont {H.~M.}\ \bibnamefont {Barbosa}}, \bibinfo {author}
  {\bibfnamefont {N.}~\bibnamefont {Marwan}}, \bibinfo {author} {\bibfnamefont
  {J.}~\bibnamefont {Marengo}}, \ and\ \bibinfo {author} {\bibfnamefont
  {J.}~\bibnamefont {Kurths}},\ }\href@noop {} {\bibfield  {journal} {\bibinfo
  {journal} {Geophysical Research Letters}\ }\textbf {\bibinfo {volume} {41}},\
  \bibinfo {pages} {7397} (\bibinfo {year} {2014}{\natexlab{b}})}\BibitemShut
  {NoStop}%
\bibitem [{\citenamefont {Boers}\ \emph
  {et~al.}(2015{\natexlab{a}})\citenamefont {Boers}, \citenamefont {Donner},
  \citenamefont {Bookhagen},\ and\ \citenamefont {Kurths}}]{Boers2015}%
  \BibitemOpen
  \bibfield  {author} {\bibinfo {author} {\bibfnamefont {N.}~\bibnamefont
  {Boers}}, \bibinfo {author} {\bibfnamefont {R.~V.}\ \bibnamefont {Donner}},
  \bibinfo {author} {\bibfnamefont {B.}~\bibnamefont {Bookhagen}}, \ and\
  \bibinfo {author} {\bibfnamefont {J.}~\bibnamefont {Kurths}},\ }\href@noop {}
  {\bibfield  {journal} {\bibinfo  {journal} {Climate Dynamics}\ }\textbf
  {\bibinfo {volume} {45}},\ \bibinfo {pages} {619} (\bibinfo {year}
  {2015}{\natexlab{a}})}\BibitemShut {NoStop}%
\bibitem [{\citenamefont {Boers}\ \emph
  {et~al.}(2015{\natexlab{b}})\citenamefont {Boers}, \citenamefont {Bookhagen},
  \citenamefont {Marengo}, \citenamefont {Marwan}, \citenamefont {von Storch},\
  and\ \citenamefont {Kurths}}]{Boers2015a}%
  \BibitemOpen
  \bibfield  {author} {\bibinfo {author} {\bibfnamefont {N.}~\bibnamefont
  {Boers}}, \bibinfo {author} {\bibfnamefont {B.}~\bibnamefont {Bookhagen}},
  \bibinfo {author} {\bibfnamefont {J.}~\bibnamefont {Marengo}}, \bibinfo
  {author} {\bibfnamefont {N.}~\bibnamefont {Marwan}}, \bibinfo {author}
  {\bibfnamefont {J.-S.}\ \bibnamefont {von Storch}}, \ and\ \bibinfo {author}
  {\bibfnamefont {J.}~\bibnamefont {Kurths}},\ }\href@noop {} {\bibfield
  {journal} {\bibinfo  {journal} {Journal of Climate}\ }\textbf {\bibinfo
  {volume} {28}},\ \bibinfo {pages} {1031} (\bibinfo {year}
  {2015}{\natexlab{b}})}\BibitemShut {NoStop}%
\bibitem [{\citenamefont {Boers}\ \emph {et~al.}(2016)\citenamefont {Boers},
  \citenamefont {Bookhagen}, \citenamefont {Marwan},\ and\ \citenamefont
  {Kurths}}]{Boers2016}%
  \BibitemOpen
  \bibfield  {author} {\bibinfo {author} {\bibfnamefont {N.}~\bibnamefont
  {Boers}}, \bibinfo {author} {\bibfnamefont {B.}~\bibnamefont {Bookhagen}},
  \bibinfo {author} {\bibfnamefont {N.}~\bibnamefont {Marwan}}, \ and\ \bibinfo
  {author} {\bibfnamefont {J.}~\bibnamefont {Kurths}},\ }\href@noop {}
  {\bibfield  {journal} {\bibinfo  {journal} {Climate Dynamics}\ }\textbf
  {\bibinfo {volume} {46}},\ \bibinfo {pages} {601} (\bibinfo {year}
  {2016})}\BibitemShut {NoStop}%
\bibitem [{\citenamefont {Stolbova}\ \emph {et~al.}(2014)\citenamefont
  {Stolbova}, \citenamefont {Martin}, \citenamefont {Bookhagen}, \citenamefont
  {Marwan},\ and\ \citenamefont {Kurths}}]{Stolbova2014}%
  \BibitemOpen
  \bibfield  {author} {\bibinfo {author} {\bibfnamefont {V.}~\bibnamefont
  {Stolbova}}, \bibinfo {author} {\bibfnamefont {P.}~\bibnamefont {Martin}},
  \bibinfo {author} {\bibfnamefont {B.}~\bibnamefont {Bookhagen}}, \bibinfo
  {author} {\bibfnamefont {N.}~\bibnamefont {Marwan}}, \ and\ \bibinfo {author}
  {\bibfnamefont {J.}~\bibnamefont {Kurths}},\ }\href {\doibase
  10.5194/npg-21-901-2014} {\bibfield  {journal} {\bibinfo  {journal}
  {Nonlinear Processes in Geophysics}\ }\textbf {\bibinfo {volume} {21}},\
  \bibinfo {pages} {901} (\bibinfo {year} {2014})}\BibitemShut {NoStop}%
\bibitem [{\citenamefont {He}\ \emph {et~al.}(2014)\citenamefont {He},
  \citenamefont {Feng}, \citenamefont {Gong}, \citenamefont {Huang},
  \citenamefont {Wu},\ and\ \citenamefont {Gong}}]{He2014}%
  \BibitemOpen
  \bibfield  {author} {\bibinfo {author} {\bibfnamefont {S.-H.}\ \bibnamefont
  {He}}, \bibinfo {author} {\bibfnamefont {T.-C.}\ \bibnamefont {Feng}},
  \bibinfo {author} {\bibfnamefont {Y.-C.}\ \bibnamefont {Gong}}, \bibinfo
  {author} {\bibfnamefont {Y.-H.}\ \bibnamefont {Huang}}, \bibinfo {author}
  {\bibfnamefont {C.-G.}\ \bibnamefont {Wu}}, \ and\ \bibinfo {author}
  {\bibfnamefont {Z.-Q.}\ \bibnamefont {Gong}},\ }\href {\doibase
  10.1088/1674-1056/23/5/059202} {\bibfield  {journal} {\bibinfo  {journal}
  {Chinese Physics B}\ }\textbf {\bibinfo {volume} {23}},\ \bibinfo {pages}
  {059202} (\bibinfo {year} {2014})}\BibitemShut {NoStop}%
\bibitem [{\citenamefont {Rehfeld}\ and\ \citenamefont
  {Kurths}(2014)}]{Rehfeld2014}%
  \BibitemOpen
  \bibfield  {author} {\bibinfo {author} {\bibfnamefont {K.}~\bibnamefont
  {Rehfeld}}\ and\ \bibinfo {author} {\bibfnamefont {J.}~\bibnamefont
  {Kurths}},\ }\href@noop {} {\bibfield  {journal} {\bibinfo  {journal}
  {Climate of the Past}\ }\textbf {\bibinfo {volume} {10}},\ \bibinfo {pages}
  {107} (\bibinfo {year} {2014})}\BibitemShut {NoStop}%
\bibitem [{\citenamefont {Rheinwalt}\ \emph {et~al.}(2012)\citenamefont
  {Rheinwalt}, \citenamefont {Marwan}, \citenamefont {Kurths}, \citenamefont
  {Werner},\ and\ \citenamefont {Gerstengarbe}}]{Rheinwalt2012}%
  \BibitemOpen
  \bibfield  {author} {\bibinfo {author} {\bibfnamefont {A.}~\bibnamefont
  {Rheinwalt}}, \bibinfo {author} {\bibfnamefont {N.}~\bibnamefont {Marwan}},
  \bibinfo {author} {\bibfnamefont {J.}~\bibnamefont {Kurths}}, \bibinfo
  {author} {\bibfnamefont {P.}~\bibnamefont {Werner}}, \ and\ \bibinfo {author}
  {\bibfnamefont {F.-W.}\ \bibnamefont {Gerstengarbe}},\ }in\ \href@noop {}
  {\emph {\bibinfo {booktitle} {High Performance Computing, Networking, Storage
  and Analysis (SCC), 2012 SC Companion:}}}\ (\bibinfo {organization} {IEEE},\
  \bibinfo {year} {2012})\ pp.\ \bibinfo {pages} {500--505}\BibitemShut
  {NoStop}%
\bibitem [{\citenamefont {Marwan}\ and\ \citenamefont
  {Kurths}(2015)}]{Marwan2015}%
  \BibitemOpen
  \bibfield  {author} {\bibinfo {author} {\bibfnamefont {N.}~\bibnamefont
  {Marwan}}\ and\ \bibinfo {author} {\bibfnamefont {J.}~\bibnamefont
  {Kurths}},\ }\href@noop {} {\bibfield  {journal} {\bibinfo  {journal} {Chaos:
  An Interdisciplinary Journal of Nonlinear Science}\ }\textbf {\bibinfo
  {volume} {25}},\ \bibinfo {pages} {097609} (\bibinfo {year}
  {2015})}\BibitemShut {NoStop}%
\bibitem [{\citenamefont {Donner}\ \emph {et~al.}(2017)\citenamefont {Donner},
  \citenamefont {Wiedermann},\ and\ \citenamefont {Donges}}]{Donner2017}%
  \BibitemOpen
  \bibfield  {author} {\bibinfo {author} {\bibfnamefont {R.~V.}\ \bibnamefont
  {Donner}}, \bibinfo {author} {\bibfnamefont {M.}~\bibnamefont {Wiedermann}},
  \ and\ \bibinfo {author} {\bibfnamefont {J.~F.}\ \bibnamefont {Donges}},\
  }in\ \href@noop {} {\emph {\bibinfo {booktitle} {Nonlinear and Stochastic
  Climate Dynamics}}},\ \bibinfo {editor} {edited by\ \bibinfo {editor}
  {\bibfnamefont {C.}~\bibnamefont {Franzke}}\ and\ \bibinfo {editor}
  {\bibfnamefont {T.}~\bibnamefont {O'Kane}}}\ (\bibinfo  {publisher}
  {Cambridge University Press},\ \bibinfo {address} {Cambridge},\ \bibinfo
  {year} {2017})\ \bibinfo {edition} {1st}\ ed.,\ pp.\ \bibinfo {pages}
  {159--183}\BibitemShut {NoStop}%
\bibitem [{\citenamefont {Dijkstra}\ \emph {et~al.}(2019)\citenamefont
  {Dijkstra}, \citenamefont {Hern\'andez-Garc\'ia}, \citenamefont {Masoller},\
  and\ \citenamefont {Barreiro}}]{Dijkstra2019}%
  \BibitemOpen
  \bibfield  {author} {\bibinfo {author} {\bibfnamefont {H.~A.}\ \bibnamefont
  {Dijkstra}}, \bibinfo {author} {\bibfnamefont {E.}~\bibnamefont
  {Hern\'andez-Garc\'ia}}, \bibinfo {author} {\bibfnamefont {C.}~\bibnamefont
  {Masoller}}, \ and\ \bibinfo {author} {\bibfnamefont {M.}~\bibnamefont
  {Barreiro}},\ }\href@noop {} {\emph {\bibinfo {title} {Networks in
  climate}}}\ (\bibinfo  {publisher} {Cambridge University Press},\ \bibinfo
  {address} {Cambridge},\ \bibinfo {year} {2019})\BibitemShut {NoStop}%
\bibitem [{\citenamefont {Schleussner}\ \emph {et~al.}(2016)\citenamefont
  {Schleussner}, \citenamefont {Donges}, \citenamefont {Donner},\ and\
  \citenamefont {Schellnhuber}}]{Schleussner2016}%
  \BibitemOpen
  \bibfield  {author} {\bibinfo {author} {\bibfnamefont {C.-F.}\ \bibnamefont
  {Schleussner}}, \bibinfo {author} {\bibfnamefont {J.~F.}\ \bibnamefont
  {Donges}}, \bibinfo {author} {\bibfnamefont {R.~V.}\ \bibnamefont {Donner}},
  \ and\ \bibinfo {author} {\bibfnamefont {H.~J.}\ \bibnamefont
  {Schellnhuber}},\ }\href {\doibase 10.1073/pnas.1601611113} {\bibfield
  {journal} {\bibinfo  {journal} {Proceedings of the National Academy of
  Sciences}\ }\textbf {\bibinfo {volume} {113}},\ \bibinfo {pages} {9216}
  (\bibinfo {year} {2016})}\BibitemShut {NoStop}%
\bibitem [{\citenamefont {Siegmund}\ \emph
  {et~al.}(2016{\natexlab{a}})\citenamefont {Siegmund}, \citenamefont
  {Sanders}, \citenamefont {Heinrich}, \citenamefont {van~der Maaten},
  \citenamefont {Simard}, \citenamefont {Helle},\ and\ \citenamefont
  {Donner}}]{Siegmund2016}%
  \BibitemOpen
  \bibfield  {author} {\bibinfo {author} {\bibfnamefont {J.~F.}\ \bibnamefont
  {Siegmund}}, \bibinfo {author} {\bibfnamefont {T.~G.~M.}\ \bibnamefont
  {Sanders}}, \bibinfo {author} {\bibfnamefont {I.}~\bibnamefont {Heinrich}},
  \bibinfo {author} {\bibfnamefont {E.}~\bibnamefont {van~der Maaten}},
  \bibinfo {author} {\bibfnamefont {S.}~\bibnamefont {Simard}}, \bibinfo
  {author} {\bibfnamefont {G.}~\bibnamefont {Helle}}, \ and\ \bibinfo {author}
  {\bibfnamefont {R.~V.}\ \bibnamefont {Donner}},\ }\href {\doibase
  10.3389/fpls.2016.00733} {\bibfield  {journal} {\bibinfo  {journal}
  {Frontiers in Plant Science}\ }\textbf {\bibinfo {volume} {7}},\ \bibinfo
  {pages} {733} (\bibinfo {year} {2016}{\natexlab{a}})}\BibitemShut {NoStop}%
\bibitem [{\citenamefont {Siegmund}\ \emph
  {et~al.}(2016{\natexlab{b}})\citenamefont {Siegmund}, \citenamefont
  {Wiedermann}, \citenamefont {Donges},\ and\ \citenamefont
  {Donner}}]{Siegmund2016a}%
  \BibitemOpen
  \bibfield  {author} {\bibinfo {author} {\bibfnamefont {J.~F.}\ \bibnamefont
  {Siegmund}}, \bibinfo {author} {\bibfnamefont {M.}~\bibnamefont
  {Wiedermann}}, \bibinfo {author} {\bibfnamefont {J.~F.}\ \bibnamefont
  {Donges}}, \ and\ \bibinfo {author} {\bibfnamefont {R.~V.}\ \bibnamefont
  {Donner}},\ }\href@noop {} {\bibfield  {journal} {\bibinfo  {journal}
  {Biogeosciences}\ }\textbf {\bibinfo {volume} {13}},\ \bibinfo {pages} {5541}
  (\bibinfo {year} {2016}{\natexlab{b}})}\BibitemShut {NoStop}%
\bibitem [{\citenamefont {Siegmund}\ \emph {et~al.}(2017)\citenamefont
  {Siegmund}, \citenamefont {Siegmund},\ and\ \citenamefont
  {Donner}}]{Siegmund2017}%
  \BibitemOpen
  \bibfield  {author} {\bibinfo {author} {\bibfnamefont {J.~F.}\ \bibnamefont
  {Siegmund}}, \bibinfo {author} {\bibfnamefont {N.}~\bibnamefont {Siegmund}},
  \ and\ \bibinfo {author} {\bibfnamefont {R.~V.}\ \bibnamefont {Donner}},\
  }\href {\doibase 10.1016/j.cageo.2016.10.004} {\bibfield  {journal} {\bibinfo
   {journal} {Computers \& Geosciences}\ }\textbf {\bibinfo {volume} {98}},\
  \bibinfo {pages} {64 } (\bibinfo {year} {2017})}\BibitemShut {NoStop}%
\bibitem [{\citenamefont {Rammig}\ \emph {et~al.}(2015)\citenamefont {Rammig},
  \citenamefont {Wiedermann}, \citenamefont {Donges}, \citenamefont {Babst},
  \citenamefont {von Bloh}, \citenamefont {Frank}, \citenamefont {Thonicke},\
  and\ \citenamefont {Mahecha}}]{Rammig2015}%
  \BibitemOpen
  \bibfield  {author} {\bibinfo {author} {\bibfnamefont {A.}~\bibnamefont
  {Rammig}}, \bibinfo {author} {\bibfnamefont {M.}~\bibnamefont {Wiedermann}},
  \bibinfo {author} {\bibfnamefont {J.~F.}\ \bibnamefont {Donges}}, \bibinfo
  {author} {\bibfnamefont {F.}~\bibnamefont {Babst}}, \bibinfo {author}
  {\bibfnamefont {W.}~\bibnamefont {von Bloh}}, \bibinfo {author}
  {\bibfnamefont {D.}~\bibnamefont {Frank}}, \bibinfo {author} {\bibfnamefont
  {K.}~\bibnamefont {Thonicke}}, \ and\ \bibinfo {author} {\bibfnamefont
  {M.~D.}\ \bibnamefont {Mahecha}},\ }\href {\doibase 10.5194/bg-12-373-2015}
  {\bibfield  {journal} {\bibinfo  {journal} {Biogeosciences}\ }\textbf
  {\bibinfo {volume} {12}},\ \bibinfo {pages} {373} (\bibinfo {year}
  {2015})}\BibitemShut {NoStop}%
\bibitem [{\citenamefont {Baumbach}\ \emph {et~al.}(2017)\citenamefont
  {Baumbach}, \citenamefont {Siegmund}, \citenamefont {Mittermeier},\ and\
  \citenamefont {Donner}}]{Baumbach2017}%
  \BibitemOpen
  \bibfield  {author} {\bibinfo {author} {\bibfnamefont {L.}~\bibnamefont
  {Baumbach}}, \bibinfo {author} {\bibfnamefont {J.~F.}\ \bibnamefont
  {Siegmund}}, \bibinfo {author} {\bibfnamefont {M.}~\bibnamefont
  {Mittermeier}}, \ and\ \bibinfo {author} {\bibfnamefont {R.~V.}\ \bibnamefont
  {Donner}},\ }\href@noop {} {\bibfield  {journal} {\bibinfo  {journal}
  {Biogeosciences}\ }\textbf {\bibinfo {volume} {14}},\ \bibinfo {pages} {4891}
  (\bibinfo {year} {2017})}\BibitemShut {NoStop}%
\bibitem [{\citenamefont {Wiedermann}\ \emph {et~al.}(2017)\citenamefont
  {Wiedermann}, \citenamefont {Siegmund}, \citenamefont {Donges}, \citenamefont
  {Kurths},\ and\ \citenamefont {Donner}}]{Wiedermann2017}%
  \BibitemOpen
  \bibfield  {author} {\bibinfo {author} {\bibfnamefont {M.}~\bibnamefont
  {Wiedermann}}, \bibinfo {author} {\bibfnamefont {J.~F.}\ \bibnamefont
  {Siegmund}}, \bibinfo {author} {\bibfnamefont {J.~F.}\ \bibnamefont
  {Donges}}, \bibinfo {author} {\bibfnamefont {J.}~\bibnamefont {Kurths}}, \
  and\ \bibinfo {author} {\bibfnamefont {R.~V.}\ \bibnamefont {Donner}},\
  }\href@noop {} {\enquote {\bibinfo {title} {Differential imprints of distinct
  enso flavors in global extreme precipitation patterns},}\ } (\bibinfo {year}
  {2017}),\ \Eprint {http://arxiv.org/abs/1702.00218} {arXiv:1702.00218
  [physics.ao-ph]} \BibitemShut {NoStop}%
\bibitem [{\citenamefont {Sippel}\ \emph {et~al.}(2017)\citenamefont {Sippel},
  \citenamefont {Zscheischler}, \citenamefont {Mahecha}, \citenamefont {Orth},
  \citenamefont {Reichstein}, \citenamefont {Vogel},\ and\ \citenamefont
  {Seneviratne}}]{Sippel2017}%
  \BibitemOpen
  \bibfield  {author} {\bibinfo {author} {\bibfnamefont {S.}~\bibnamefont
  {Sippel}}, \bibinfo {author} {\bibfnamefont {J.}~\bibnamefont
  {Zscheischler}}, \bibinfo {author} {\bibfnamefont {M.~D.}\ \bibnamefont
  {Mahecha}}, \bibinfo {author} {\bibfnamefont {R.}~\bibnamefont {Orth}},
  \bibinfo {author} {\bibfnamefont {M.}~\bibnamefont {Reichstein}}, \bibinfo
  {author} {\bibfnamefont {M.}~\bibnamefont {Vogel}}, \ and\ \bibinfo {author}
  {\bibfnamefont {S.~I.}\ \bibnamefont {Seneviratne}},\ }\href@noop {}
  {\bibfield  {journal} {\bibinfo  {journal} {Earth System Dynamics}\ }\textbf
  {\bibinfo {volume} {8}},\ \bibinfo {pages} {387} (\bibinfo {year}
  {2017})}\BibitemShut {NoStop}%
\bibitem [{\citenamefont {Sarlis}(2018)}]{Sarlis2018}%
  \BibitemOpen
  \bibfield  {author} {\bibinfo {author} {\bibfnamefont {N.}~\bibnamefont
  {Sarlis}},\ }\href@noop {} {\bibfield  {journal} {\bibinfo  {journal}
  {Entropy}\ }\textbf {\bibinfo {volume} {20}},\ \bibinfo {pages} {561}
  (\bibinfo {year} {2018})}\BibitemShut {NoStop}%
\bibitem [{\citenamefont {Wolf}\ \emph {et~al.}(2020)\citenamefont {Wolf},
  \citenamefont {Bauer}, \citenamefont {Boers},\ and\ \citenamefont
  {Donner}}]{Wolf2020}%
  \BibitemOpen
  \bibfield  {author} {\bibinfo {author} {\bibfnamefont {F.}~\bibnamefont
  {Wolf}}, \bibinfo {author} {\bibfnamefont {J.}~\bibnamefont {Bauer}},
  \bibinfo {author} {\bibfnamefont {N.}~\bibnamefont {Boers}}, \ and\ \bibinfo
  {author} {\bibfnamefont {R.~V.}\ \bibnamefont {Donner}},\ }\href {\doibase
  10.1063/1.5134012} {\bibfield  {journal} {\bibinfo  {journal} {Chaos}\
  }\textbf {\bibinfo {volume} {30}},\ \bibinfo {pages} {033102} (\bibinfo
  {year} {2020})}\BibitemShut {NoStop}%
\bibitem [{\citenamefont {Kreuz}\ \emph {et~al.}(2015)\citenamefont {Kreuz},
  \citenamefont {Mulansky},\ and\ \citenamefont {Bozanic}}]{Kreuz2015}%
  \BibitemOpen
  \bibfield  {author} {\bibinfo {author} {\bibfnamefont {T.}~\bibnamefont
  {Kreuz}}, \bibinfo {author} {\bibfnamefont {M.}~\bibnamefont {Mulansky}}, \
  and\ \bibinfo {author} {\bibfnamefont {N.}~\bibnamefont {Bozanic}},\
  }\href@noop {} {\bibfield  {journal} {\bibinfo  {journal} {Journal of
  Neurophysiology}\ }\textbf {\bibinfo {volume} {113}},\ \bibinfo {pages}
  {3432} (\bibinfo {year} {2015})}\BibitemShut {NoStop}%
\bibitem [{\citenamefont {Mulansky}\ \emph {et~al.}(2015)\citenamefont
  {Mulansky}, \citenamefont {Bozanic}, \citenamefont {Sburlea},\ and\
  \citenamefont {Kreuz}}]{Mulansky2015}%
  \BibitemOpen
  \bibfield  {author} {\bibinfo {author} {\bibfnamefont {M.}~\bibnamefont
  {Mulansky}}, \bibinfo {author} {\bibfnamefont {N.}~\bibnamefont {Bozanic}},
  \bibinfo {author} {\bibfnamefont {A.}~\bibnamefont {Sburlea}}, \ and\
  \bibinfo {author} {\bibfnamefont {T.}~\bibnamefont {Kreuz}},\ }in\ \href
  {\doibase 10.1109/ebccsp.2015.7300693} {\emph {\bibinfo {booktitle} {2015
  International Conference on Event-based Control, Communication, and Signal
  Processing ({EBCCSP})}}}\ (\bibinfo  {publisher} {{IEEE}},\ \bibinfo {year}
  {2015})\ pp.\ \bibinfo {pages} {1--8}\BibitemShut {NoStop}%
\bibitem [{\citenamefont {Kropp}\ and\ \citenamefont
  {Schellnhuber}(2011)}]{Kropp2011}%
  \BibitemOpen
  \bibinfo {editor} {\bibfnamefont {J.}~\bibnamefont {Kropp}}\ and\ \bibinfo
  {editor} {\bibfnamefont {H.-J.}\ \bibnamefont {Schellnhuber}},\ eds.,\
  \href@noop {} {\emph {\bibinfo {title} {In Extremis - Disruptive Events and
  Trends in Climate and Hydrology}}}\ (\bibinfo  {publisher} {Springer},\
  \bibinfo {address} {Berlin},\ \bibinfo {year} {2011})\BibitemShut {NoStop}%
\bibitem [{\citenamefont {Costa}\ \emph {et~al.}(2007)\citenamefont {Costa},
  \citenamefont {Rodrigues}, \citenamefont {Travieso},\ and\ \citenamefont
  {Boas}}]{Costa2007}%
  \BibitemOpen
  \bibfield  {author} {\bibinfo {author} {\bibfnamefont {L.~d.~F.}\
  \bibnamefont {Costa}}, \bibinfo {author} {\bibfnamefont {F.~A.}\ \bibnamefont
  {Rodrigues}}, \bibinfo {author} {\bibfnamefont {G.}~\bibnamefont {Travieso}},
  \ and\ \bibinfo {author} {\bibfnamefont {P.~R.~V.}\ \bibnamefont {Boas}},\
  }\href {\doibase 10.1080/00018730601170527} {\bibfield  {journal} {\bibinfo
  {journal} {Advances in Physics}\ }\textbf {\bibinfo {volume} {56}},\ \bibinfo
  {pages} {167} (\bibinfo {year} {2007})}\BibitemShut {NoStop}%
\bibitem [{\citenamefont {{Goddard Earth Sciences Data and Information Services
  Center}}(2016)}]{GESDISC2016}%
  \BibitemOpen
  \bibfield  {author} {\bibinfo {author} {\bibnamefont {{Goddard Earth Sciences
  Data and Information Services Center}}},\ }\href
  {http://disc.gsfc.nasa.gov/datacollection/TRMM_3B42_Daily_7.html} {\emph
  {\bibinfo {title} {TRMM (TMPA) Precipitation L3 1 day 0.25 degree x 0.25
  degree V7}}},\ \bibinfo {organization} {Goddard Earth Sciences Data and
  Information Services Center (GES DISC)} (\bibinfo {year} {2016})\BibitemShut
  {NoStop}%
\bibitem [{Note1()}]{Note1}%
  \BibitemOpen
  \bibinfo {note} {The rat EEG data may be downloaded from \protect \url
  {https://www2.le.ac.uk/centres/csn/software} as dataset 4.}\BibitemShut
  {Stop}%
\bibitem [{\citenamefont {van Luijtelaar}(1997)}]{Luijtelaar1997}%
  \BibitemOpen
  \bibfield  {author} {\bibinfo {author} {\bibfnamefont {G.}~\bibnamefont {van
  Luijtelaar}},\ }\href@noop {} {\emph {\bibinfo {title} {The WAG/Rij Rat Model
  of Absence Epilepsy: Ten Years of Research: a Compilation of Papers}}}\
  (\bibinfo  {publisher} {Nijmegen University Press},\ \bibinfo {year}
  {1997})\BibitemShut {NoStop}%
\bibitem [{\citenamefont {Quian~Quiroga}\ \emph
  {et~al.}(2002{\natexlab{b}})\citenamefont {Quian~Quiroga}, \citenamefont
  {Kraskov}, \citenamefont {Kreuz},\ and\ \citenamefont
  {Grassberger}}]{QuianQuiroga2002a}%
  \BibitemOpen
  \bibfield  {author} {\bibinfo {author} {\bibfnamefont {R.}~\bibnamefont
  {Quian~Quiroga}}, \bibinfo {author} {\bibfnamefont {A.}~\bibnamefont
  {Kraskov}}, \bibinfo {author} {\bibfnamefont {T.}~\bibnamefont {Kreuz}}, \
  and\ \bibinfo {author} {\bibfnamefont {P.}~\bibnamefont {Grassberger}},\
  }\href {\doibase 10.1103/PhysRevE.65.041903} {\bibfield  {journal} {\bibinfo
  {journal} {Phys. Rev. E}\ }\textbf {\bibinfo {volume} {65}},\ \bibinfo
  {pages} {041903} (\bibinfo {year} {2002}{\natexlab{b}})}\BibitemShut
  {NoStop}%
\bibitem [{\citenamefont {Ferreira}\ \emph {et~al.}(2019)\citenamefont
  {Ferreira}, \citenamefont {Ferreira}, \citenamefont {Gava}, \citenamefont
  {Zhao},\ and\ \citenamefont {Macau}}]{Ferreira2019}%
  \BibitemOpen
  \bibfield  {author} {\bibinfo {author} {\bibfnamefont {L.~N.}\ \bibnamefont
  {Ferreira}}, \bibinfo {author} {\bibfnamefont {N.~C.~R.}\ \bibnamefont
  {Ferreira}}, \bibinfo {author} {\bibfnamefont {M.~L. L.~M.}\ \bibnamefont
  {Gava}}, \bibinfo {author} {\bibfnamefont {L.}~\bibnamefont {Zhao}}, \ and\
  \bibinfo {author} {\bibfnamefont {E.~E.~N.}\ \bibnamefont {Macau}},\
  }\href@noop {} {\enquote {\bibinfo {title} {The influence of time series
  distance functions on climate networks},}\ } (\bibinfo {year} {2019}),\
  \Eprint {http://arxiv.org/abs/1902.03298} {arXiv:1902.03298
  [physics.data-an]} \BibitemShut {NoStop}%
\bibitem [{\citenamefont {Hassanibesheli}\ and\ \citenamefont
  {Donner}(2019)}]{Hassanibesheli2019}%
  \BibitemOpen
  \bibfield  {author} {\bibinfo {author} {\bibfnamefont {F.}~\bibnamefont
  {Hassanibesheli}}\ and\ \bibinfo {author} {\bibfnamefont {R.~V.}\
  \bibnamefont {Donner}},\ }\href@noop {} {\bibfield  {journal} {\bibinfo
  {journal} {Chaos}\ }\textbf {\bibinfo {volume} {29}},\ \bibinfo {pages}
  {083125} (\bibinfo {year} {2019})}\BibitemShut {NoStop}%
\bibitem [{\citenamefont {Donges}\ \emph {et~al.}(2015)\citenamefont {Donges},
  \citenamefont {Heitzig}, \citenamefont {Beronov}, \citenamefont {Wiedermann},
  \citenamefont {Runge}, \citenamefont {Feng}, \citenamefont {Tupikina},
  \citenamefont {Stolbova}, \citenamefont {Donner}, \citenamefont {Marwan},
  \citenamefont {Dijkstra},\ and\ \citenamefont {Kurths}}]{Donges2015}%
  \BibitemOpen
  \bibfield  {author} {\bibinfo {author} {\bibfnamefont {J.~F.}\ \bibnamefont
  {Donges}}, \bibinfo {author} {\bibfnamefont {J.}~\bibnamefont {Heitzig}},
  \bibinfo {author} {\bibfnamefont {B.}~\bibnamefont {Beronov}}, \bibinfo
  {author} {\bibfnamefont {M.}~\bibnamefont {Wiedermann}}, \bibinfo {author}
  {\bibfnamefont {J.}~\bibnamefont {Runge}}, \bibinfo {author} {\bibfnamefont
  {Q.~Y.}\ \bibnamefont {Feng}}, \bibinfo {author} {\bibfnamefont
  {L.}~\bibnamefont {Tupikina}}, \bibinfo {author} {\bibfnamefont
  {V.}~\bibnamefont {Stolbova}}, \bibinfo {author} {\bibfnamefont {R.~V.}\
  \bibnamefont {Donner}}, \bibinfo {author} {\bibfnamefont {N.}~\bibnamefont
  {Marwan}}, \bibinfo {author} {\bibfnamefont {H.~A.}\ \bibnamefont
  {Dijkstra}}, \ and\ \bibinfo {author} {\bibfnamefont {J.}~\bibnamefont
  {Kurths}},\ }\href@noop {} {\bibfield  {journal} {\bibinfo  {journal}
  {Chaos}\ }\textbf {\bibinfo {volume} {25}} (\bibinfo {year}
  {2015})}\BibitemShut {NoStop}%
\end{thebibliography}%

\end{document}